\documentclass[10pt,aps,pra,twocolumn,superscriptaddress,floatfix,nofootinbib]{revtex4-2}
\usepackage{caption}
\usepackage{lmodern}
\usepackage{silence}
\pdfoutput=1
 
\usepackage{physics}
\usepackage{amsmath}
\usepackage{amssymb}
\usepackage{amsthm}
\usepackage{graphicx}
\usepackage{hyperref}
\usepackage{mathtools}
\usepackage{ragged2e}
\usepackage{microtype}
\usepackage{subcaption}
\usepackage[all]{nowidow}
\usepackage{thm-restate}
\usepackage{makecell}
\usepackage{thmtools}
\usepackage[capitalise]{cleveref}
\graphicspath{{graphics/}}
\usepackage{amscd,amsfonts,mathrsfs,braket,bbm,enumerate,bm}
\usepackage[dvipsnames]{xcolor}
\newcommand{\R}{\mathbb{R}}
\usepackage{etoolbox}
\usepackage{tikz}
\usepackage{graphicx}
\usetikzlibrary{arrows.meta,calc,positioning,backgrounds,decorations.pathreplacing}
\usetikzlibrary{quantikz2}
\usepackage{algorithm}
\usepackage{algorithmic}
\usepackage{tikz}
\usepackage{hhline}
\usepackage{colortbl}
\usepackage{booktabs}
\usepackage{fontawesome}
\usepackage{adjustbox}
\usepackage{etoolbox}

\usepackage{braket}
\usepackage{sourcesanspro}
\allowdisplaybreaks

\newtheorem{theorem}{Theorem}

\theoremstyle{definition}
\newtheorem{definition}{Definition}

\usepackage{tikz}
\usetikzlibrary{arrows.meta,positioning,fit,calc,decorations.pathmorphing,backgrounds}

\definecolor{linkcolor}{HTML}{8B1E3F}
\definecolor{citecolor}{HTML}{1F4E79}
\definecolor{urlcolor}{HTML}{5E3C99}

\hypersetup{colorlinks=true,
    citecolor=citecolor,
    linkcolor=linkcolor,
    urlcolor=urlcolor
}

\addtocounter{problem}{-1}

\begin{document}
\title{Quantum Topological Data Encoding}

\author{Adam Wesołowski}
\affiliation{Royal Holloway University of London, Department of Computer Science, UK}

\author{Dimitrios Thanos}
\affiliation{Leiden Institute of Advanced Computer Science (LIACS), Leiden University, Leiden, The Netherlands}

\author{Daniel Leykam}
\affiliation{Science, Mathematics and Technology Cluster, Singapore University of Technology and Design, Singapore}

\author{Lirandë Pira}
\email{lpira@nus.edu.sg}
\affiliation{Centre for Quantum Technologies, National University of Singapore, Singapore}

\date{\today}

\begin{abstract}
Many datasets encountered across a wide range of domains possess rich geometric and topological structure that is difficult to capture using conventional vector-based representations. Quantum machine learning offers the possibility of processing high-dimensional data in Hilbert spaces, but its practical success depends critically on how classical data is encoded into quantum states. We introduce \emph{quantum topological data encoding} (QTDE), a general framework for encoding topological information into quantum states via topology-driven quantum evolution. Our method generalises an existing topology-driven quantum encoding framework to higher-dimensional data. We test the proposed method on clique-complexes classification tasks, and provide preliminary evidence that topology-driven quantum representations can capture discriminative information beyond that available through direct comparisons of classical topological descriptors. The proposed quantum representations consistently outperform a baseline based on direct comparisons of the combinatorial Laplacians describing the underlying topological structure. We indicate several areas of application where the framework can be used to provide a more efficient and reliable data representation. 
\end{abstract}

\maketitle

\section{Introduction}

\begin{figure*}[t!]
    \centering
    \makebox[\textwidth][c]{%
        \resizebox{0.99\textwidth}{!}{%
            \input{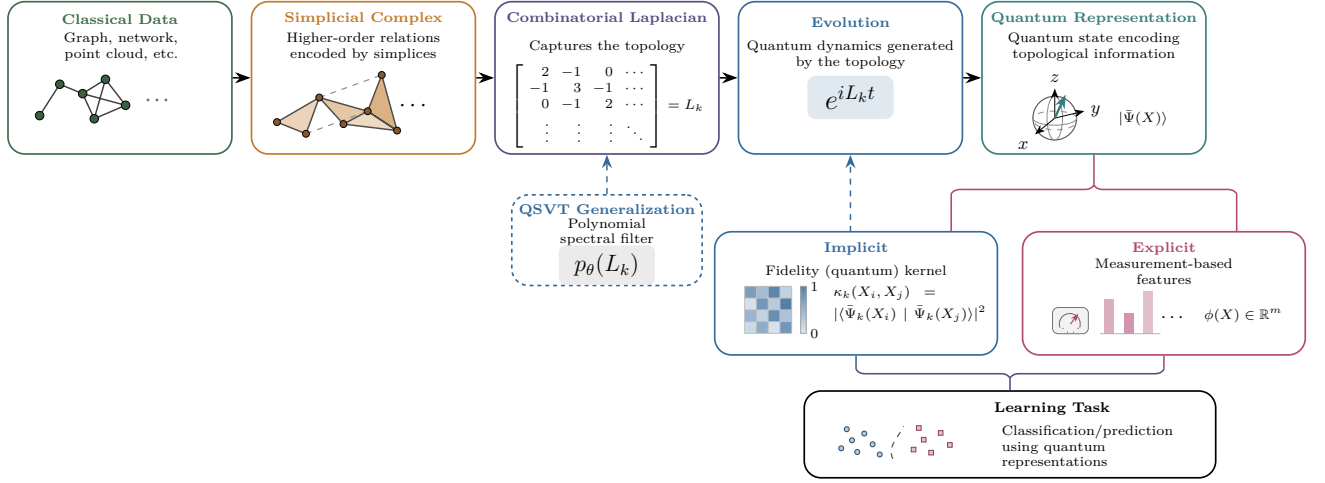}%
        }%
    }
    \caption{\justifying Overview of the \emph{Quantum Topological Data Encoding} (QTDE) framework.
    Classical relational data are lifted to a simplicial complex, encoded by the combinatorial Laplacian $L_k$, and mapped to quantum representations through topology-driven evolution or spectral filtering implemented via quantum singular value transformation  (QSVT). The resulting quantum representation can be accessed either implicitly through a fidelity-based quantum kernel or explicitly through features derived through measurement. Both representations provide topological feature maps that can be used in downstream machine learning tasks such as classification and prediction.
    }
    \label{fig:main_figure}
\end{figure*}

The ability to extract meaningful information from complex data lies at the heart of modern science and engineering. Across disciplines ranging from physics and chemistry to biology and the social sciences, advances increasingly depend on our ability to construct representations of data that expose the underlying structure relevant to a given task. 
Machine learning has emerged as a powerful framework for automatically discovering useful representations from data~\cite{bishop2006pattern,murphy2012machine,goodfellow2016deeplearning}. At the same time, heightened interest in quantum computing~\cite{manin2007mathematics,benioff1980computer,feynman1982simulating} has motivated the development of quantum machine learning, which seeks to exploit quantum mechanical systems as computational models for processing and learning from information~\cite{aimeur2006machinelearning,dunjko2016quantumenhanced,dunjko2018machinelearning,beer2020training,chang2025primerquantummachinelearning}. One of the main challenges in quantum machine learning is how to encode classical data into quantum states in a way that is both informative and efficient~\cite{schuld2021machine,LaRose_2020,lloyd2020quantumembeddingsmachinelearning,perezsalinas2020datareuploading,hur2024neuralquantumembedding}. The choice of encoding largely determines which features of the data become accessible to subsequent quantum processing and therefore plays a crucial role in the success of quantum learning algorithms~\cite{shin2023exponential,caro2021encodingdependent,le2025optimizing,zang2025benchmarkingdataencodingmethods}.

A common perspective in quantum machine learning is to regard quantum circuits as feature maps into high-dimensional Hilbert spaces~\cite{Havl19,Schuld_2019,jager2026quantumfeaturemap,marshall2023highdimensional}, where learning is performed through inner products between quantum states. This viewpoint underlies quantum kernel methods and highlights that the effectiveness of a quantum learning algorithm depends fundamentally on the chosen embedding of classical data into quantum states~\cite{schuld2021supervisedquantummachinelearning,tanner2026nonvariationalsupervisedquantumkernel}.

Many existing quantum encoding strategies treat data as vectors and map their coordinates directly into amplitudes, phases, or circuit parameters. While such approaches are often straightforward to implement, they typically focus on the individual attributes of data points and may fail to capture more intricate structural information. In many applications, however, the relationships between data points are as important as the points themselves.

Topological data analysis provides a framework for capturing such structure by constructing simplicial complexes that encode higher-order interactions and by studying associated operators such as the combinatorial Laplacian~\cite{carlsson2009topology,edelsbrunner2010computational,carlsson2020topological,chazal2021introduction,leykam2023topological}. These objects reveal topological properties of a dataset that are not accessible through pairwise relationships alone. Combinatorial Laplacians can be used as an access model that allows one to efficiently extract topological information from data~\cite{Scali_2024}, which can be used to detect clusters, holes, and anomalies~\cite{carlsson2021topological,10.1145/2488608.2488720}. Examples where topological data analysis has been shown to be an effective tool for describing complex interactions include social and biological networks~\cite{topaz2015topological}, molecular systems~\cite{wu2023novel}, and social behaviour analysis such as how cohesive or fragmented voting behavior is~\cite{lum2013extractingwithtda}. Importantly, topological data analysis has been widely used and considered an important tool in cancer research~\cite{rabadan2020identification,nicolau2011topology} and medical imaging~\cite{singh2023topological}. In these datasets the meaningful information is often encoded in higher-order interactions and global structural patterns rather than in isolated features alone.


Quantum analogues of topological data analysis have been of interest since the early days of quantum machine learning~\cite{lloyd2016quantum}. Many works have studied the quantum computational complexity of estimating the topological invariants of the high dimensional spaces, such as the Betti number or persistent Betti number~\cite{lloyd2016quantum, hayakawa2022quantumalgorithm,berry2024analyzingprospects}. Further recent quantum approaches combining topology and machine learning have largely focused on computing topological descriptors such as Betti numbers or persistence summaries using quantum algorithms and employing them within kernel methods~\cite{incudini2023higherorder}. Over the years, the area of quantum topological data analysis has been considered as a fertile ground to search for exponential quantum advantage~\cite{Gyurik2022towardsquantum,ubaru2021quantumtopologicaldataanalysis,gyurik2026provable,mcardle2026streamlinedquantum,nghiem2025quantumtopologicaldataanalysis}. It is worth noting that beyond the analysis of classical data, topological methods have also been employed to characterize quantum states themselves, including multipartite entanglement through persistent homology~\cite{mengoni2019persistent}. These developments also motivate a complementary use of topological operators, not only as objects from which to extract invariants, but as generators of quantum representations in their own right.

The choice of encoding classical data into quantum states, for example amplitude encoding or basis encoding, directly determines the
circuit depth required for state preparation, the
expressivity of the resulting model, and the inductive
biases that the quantum algorithm inherits with respect to the underlying data~\cite{schuld2021effect}. 
In practice, the effectiveness of a quantum encoding is inherently problem dependent. Existing approaches often prioritize efficient state preparation or generic vector embeddings, but may fail to exploit the geometric or topological structure present in relational data.

In many relational learning problems, the data naturally admits a higher-order topological representation, for example through a simplicial complex. Existing approaches typically use this topology to extract descriptors or invariants \textit{after} the data has been represented. In contrast, we propose to use the topology itself to guide the construction of the quantum representation, allowing the underlying topological structure to shape the encoding from the outset.
This observation motivates the central question we address in this work:

\begin{center}
  \textit{Can the topology of a dataset directly determine the construction of a quantum representation?}
\end{center}

We introduce \emph{quantum topological data encoding} (QTDE), a framework that maps simplicial complexes to quantum states via topology-driven evolution. To this end, we propose a generalization to the method originally proposed in~\cite{PhysRevA.104.032416} for preparing quantum states representing topological data. In contrast to~\cite{incudini2023higherorder}, our approach does not compute topological invariants explicitly. Instead, the combinatorial Laplacian itself serves as the generator of a topology-driven quantum evolution, producing quantum feature maps that retain richer spectral information. This conceptual framework is depicted in~\Cref{fig:main_figure}. We detail our main contributions as follows.

Given a \textit{combinatorial Laplacian} matrix $L_k$ which describes a discrete topological structure $\mathcal{T}$ called a \textit{simplicial complex} in dimension $k$, we propose the following feature map: $\mathcal{T}\mapsto \mathcal{H}$ which maps each $k$ dimensional topological structure into a Hilbert space. Specifically, similarly to~\cite{PhysRevA.104.032416} we propose to evolve a quantum state $\ket{s}^{\otimes\binom{n}{k+1}}$ by a unitary operator defined by the topology of $\mathcal{T}$ in dimension $k$, namely $U=e^{-iL_kt}$, for some sufficiently large choice of $t$. We know $U$ is unitary since $L_k$ is Hermitian. This yields,
\begin{equation}
    e^{-iL_kt}\ket{s}^{\otimes\binom{n}{k+1}}=\ket{\Psi(\mathcal{T})}
\end{equation}
The state $\Psi(\mathcal{T})$ is a quantum state obtained by a unitary evolution process driven by the topology of $\mathcal{T}$ encoded in the matrix $L_k$. For two different topological spaces $\mathcal{T}_1$ and $\mathcal{T}_2$ and the corresponding Laplacian matrices in dimension $k$, $L_k(\mathcal{T}_1)$ and $L_k(\mathcal{T}_2)$, instead of comparing the matrices directly, we can compare the two quantum states generated by the described process. We compute the kernel between the two quantum state representations of the spaces as $K=|\braket{\Psi(\mathcal{T}_1)|\Psi(\mathcal{T}_2)}|^2$. As an alternative approach we propose a method where one evaluates the quantity $v_j=\bra{s}e^{-iL_kt_j}\ket{s}$ and forms an explicit feature vector $\vec{v}=(v_1, v_2,..., v_j)$. 

Whenever two topological spaces are of different sizes (for example two clique complexes constructed from graphs with different numbers of vertices) we perform a known technique of padding the feature vectors with zeros to match the dimensions. This corresponds to embedding the quantum state of the smaller topological space in the space that matches the size of the quantum state of the larger topological space.
In both cases the representation is generated by a unitary evolution driven by the topology of $\mathcal{T}$ in dimension $k$, which is captured by the Laplacian matrix $L_k$.

Additionally, we propose a further generalization of the prior two approaches that broadly encompasses the entire framework. Using quantum singular value transformation (QSVT)~\cite{Gily_n_2019} we can apply arbitrary polynomial $p_\theta$ to the Laplacian matrix. The advantage of using the transformed Laplacian comes from the tunable coefficients that can expose the spectral ranges of the Laplacian that are particularly relevant for separating the classes of data. 
In general, it is non-obvious whether quantum evolution driven by a dimension-dependent topology yields a meaningful representation. Nevertheless, motivated by promising results with a similar methodology for the lowest dimension (when the structure is a graph) investigated in~\cite{PhysRevA.104.032416} we propose a new extended approach allowing for general encoding of arbitrary higher dimensional data.






The rest of this manuscript is structured as follows. In~\Cref{sec:preliminaries}, we introduce the necessary preliminaries on quantum evolution kernels, simplicial complexes, and combinatorial Laplacians. \Cref{sec:qtde} presents the proposed QTDE framework, including both implicit and explicit representations, together with a polynomial spectral generalization based on quantum singular value transformation. Benchmark datasets and methodology are described in~\Cref{sec:methodology}. Numerical results are presented and analyzed in~\Cref{sec:results}. In~\Cref{sec:efficiency}, we discuss the computational complexity and implementation aspects of the proposed framework. Finally, ~\Cref{sec:conclusion} concludes the paper and outlines directions for future research.

\section{Preliminaries}\label{sec:preliminaries}
\subsection*{Notation}
\label{subsec:notation}
 
We write $[n]=\{1,\dots,n\}$ for the index set of a vertex set
$V=\{v_1,\dots,v_n\}$. Quantum states live in a finite-dimensional
complex Hilbert space $\mathbb{C}^{N}$, with the standard inner product
$\braket{\phi | \psi}$
and computational basis $\{\ket{i}\}_{i}$. A pure state is a unit vector
$\ket{\psi}\in \mathbb{C}^N$ defined up to a global phase. For a Hermitian
operator $H=H^{\dagger}$, the time evolution
$U(t)=e^{-iHt}$ is unitary for all $t\in\mathbb{R}$, and we denote by
$\{(\lambda,\ket{\phi_\lambda})\}$ its eigensystem,
$H\ket{\phi_\lambda}=\lambda\ket{\phi_\lambda}$. Matrices act as $M^{\top}$
(transpose), $M^{\dagger}$ (conjugate transpose), $\ker M$ and $\operatorname{im} M$
denote kernel and image. Throughout, $X$ denotes a classical datum
carrying relational structure (e.g.\ a graph or a network), $K(X)$ the
simplicial complex associated with it, and $L_k$ the combinatorial
Laplacian of $K(X)$ in dimension $k$.

\subsection{The Quantum Singular Value Transformation}
\label{subsec:qsvt}

Many quantum algorithms reduce to a single task: given access to a
matrix, apply a chosen function to its spectrum. Hamiltonian simulation
applies $x \mapsto e^{-ixt}$; linear-system solvers apply
$x \mapsto 1/x$; amplitude amplification applies a sign function. The
\emph{quantum singular value transformation} (QSVT) of Gily\'en, Su,
Low, and Wiebe~\cite{Gily_n_2019} realizes all of these with one primitive: given a
polynomial $p$ and a matrix $A$ encoded as a block of a unitary $U$, it
applies $p$ to the singular values of $A$.

\begin{definition}[Block-encoding]
An $(n+a)$-qubit unitary $U$ is a \emph{block-encoding} of an $n$-qubit
operator $A$ with $\lVert A\rVert \le 1$ if
\[
  A = \bigl(\langle 0|^{\otimes a}\otimes I\bigr)\,U\,
      \bigl(|0\rangle^{\otimes a}\otimes I\bigr),
\]
i.e.\ $A$ occupies the top-left block of $U$.
\end{definition}

\begin{theorem}[QSVT~\cite{Gily_n_2019}]
Let $U$ block-encode $A$ with singular value decomposition
$A = \sum_i \sigma_i\,|w_i\rangle\langle v_i|$, and let
$p \in \mathbb{R}[x]$ have degree $d$, definite parity, and
$|p(x)| \le 1$ on $[-1,1]$. Then there exist phase angles
$\Phi \in \mathbb{R}^d$ and an explicit unitary $U_\Phi$ formed by
interleaving $d$ applications of $U, U^\dagger$ with phase rotation
whose top-left block is $\sum_i p(\sigma_i)\,|w_i\rangle\langle v_i|$
(for odd $p$). The angles $\Phi$ are efficiently computable from $p$.
\end{theorem}

Transforming the spectrum by a degree-$d$ polynomial thus costs only
$d$ queries to $U$ and $O(d)$ extra gates, independent of the dimension
of $A$. 
\subsection{Simplicial Complexes}
\label{subsec:simplicial}

Manifolds and, more generally, topological spaces arise throughout
science and engineering as the natural setting for physical fields,
data distributions, and geometric shapes. A computer, however, cannot
manipulate a continuum directly: any algorithm operates on a finite
amount of data, so a space of interest must first be replaced by a
finite representation. For such a representation to be useful it must
be \emph{faithful}, meaning it should reproduce the essential topological
features of the original space (its connected components, loops,
voids, and higher-dimensional analogues) rather than merely sampling
its points. Simplicial complexes constitute a natural discretization of continuous topological spaces~\cite{carlsson2009topology}.
By gluing together elementary pieces: points, edges, triangles,
tetrahedra, and their higher-dimensional analogues, according to
simple incidence rules, simplicial complexes encode the shape of a space in a finite,
combinatorial form that is directly amenable to computation.

\subsubsection{Clique Complexes}
\label{sec: clique}

For each graph $G$ we take its \emph{clique} (or flag) complex $K(G)$: the
simplicial complex whose $k$-simplices are the $(k{+}1)$-cliques of $G$, so that
an edge is a $1$-simplex, a triangle a $2$-simplex, a tetrahedron a $3$-simplex,
and so on. The clique complex is determined entirely by the graph's edge set, namely a
set of vertices spans a simplex whenever they are pairwise adjacent which makes
it the natural higher-order structure to attach to relational data. We
use clique complexes of Erd\H{o}s--R\'enyi random graphs as the benchmark
throughout: varying the edge probability $p$ changes the density of simplices at
every dimension, giving a controlled family of topological spaces on which each
$L_k$, and hence each QTDE representation, can be evaluated independently.

\subsection{Combinatorial Laplacians}
\label{subsec:laplacian}
 
The geometry of how simplices fit together is captured by the
\emph{boundary operator} $\partial_k : C_k \to C_{k-1}$, the linear map
that sends an oriented $k$-simplex to the alternating sum of its
codimension-one faces,
\begin{equation}
  \partial_k [v_0,\dots,v_k]
    \;=\; \sum_{i=0}^{k} (-1)^{i}\,
          [v_0,\dots,\hat{v}_i,\dots,v_k],
  \label{eq:boundary}
\end{equation}
where $\hat{v}_i$ denotes omission of the $i$-th vertex. With respect to
the ordered bases of $C_k$ and $C_{k-1}$, $\partial_k$ is represented by
the \emph{incidence matrix}
$B_k\in\{-1,0,1\}^{\,n_{k-1}\times n_k}$. The defining property of a chain
complex, $\partial_{k}\partial_{k+1}=0$ (equivalently $B_kB_{k+1}=0$), states
that the boundary of a boundary is empty.
 
The \emph{$k$-th combinatorial Laplacian} is the symmetric
positive semidefinite operator on $C_k$
\begin{equation}
  L_k
    \;=\; \underbrace{B_k^{\top}B_k}_{\text{down}}
        \;+\; \underbrace{B_{k+1}B_{k+1}^{\top}}_{\text{up}}
    \;\in\; \mathbb{R}^{\,n_k\times n_k},
  \label{eq:hodge-laplacian}
\end{equation}
where the \emph{down} term $B_k^{\top}B_k$ couples $k$-simplices through
their shared $(k\!-\!1)$-faces and the \emph{up} term
$B_{k+1}B_{k+1}^{\top}$ couples them through common $(k\!+\!1)$-cofaces.
For $k=0$ the down term vanishes and \eqref{eq:hodge-laplacian} reduces to
the ordinary graph Laplacian $L_0=B_1B_1^{\top}$, so $L_k$ is a genuine
higher-order generalisation of the familiar graph operator.
 
Because each summand in \eqref{eq:hodge-laplacian} is of the form
$M^{\top}M$, the operator $L_k$ is real, symmetric, and positive
semidefinite; its eigenvalues are nonnegative and its eigenvectors form
an orthonormal basis of $C_k$. Its null space is the space of
\emph{harmonic $k$-chains}, and by the discrete Hodge theorem it is
isomorphic to the $k$-th real homology of $K$,
\begin{equation}
  \ker L_k \;\cong\; H_k(K;\mathbb{R}),
  \qquad
  \dim\ker L_k \;=\; \beta_k,
  \label{eq:hodge-betti}
\end{equation}
the $k$-th Betti number, which counts the independent $k$-dimensional
``holes'' of $K$ (connected components for $k=0$, loops for $k=1$, voids
for $k=2$, and so on). The full spectrum of $L_k$ thus refines the purely
topological information in \eqref{eq:hodge-betti} with metric, scale-like
data about the complex.
 
Two features of $L_k$ make it well suited to a quantum encoding. First,
being real symmetric, it is Hermitian, so the time evolution
\begin{equation}
  U_k(t) \;=\; e^{\,i L_k t}
  \label{eq:evolution}
\end{equation}
is unitary for every $t\in\R$ and can in principle be implemented on a
quantum device. Second, $L_k$ is sparse: the number of nonzero entries
per row is bounded by the number of faces and cofaces a single
$k$-simplex can have, which enables efficient Hamiltonian-simulation
techniques (discussed in~\Cref{sec:efficiency}). The operator
\eqref{eq:hodge-laplacian} therefore packages the higher-order topology of
$X$ into a Hermitian generator, and the unitary \eqref{eq:evolution} is the
topology-driven dynamics that the QTDE framework of~\Cref{sec:qtde} turns into a learnable quantum representation.

\subsection{Quantum Evolution Kernel}
\label{subsec:qek}
 
Kernel methods classify data through a symmetric, positive
semidefinite similarity function $K(\,\cdot\,,\cdot\,)$ that implicitly
embeds each datum into a (possibly high-dimensional) feature space, so
that learning reduces to inner-product comparisons rather than to
explicit coordinates~\cite{bishop2006pattern}. The \emph{quantum evolution
kernel} (QEK) introduced originally in~\cite{PhysRevA.104.032416} describes such an embedding for graph structures through
Hamiltonian dynamics: a graph $G$ is represented by a
Hermitian operator $H(X)$, a fixed reference state $\ket{x}$ is evolved
under it, and the resulting state
\begin{equation}
  \ket{\Psi(X)} \;=\; e^{-iH(X)\,t}\,\ket{x}
  \label{eq:qek-state}
\end{equation}
serves as the quantum feature representation of the graph.

\section{Quantum Topological Data Encoding (QTDE)}\label{sec:qtde}


\subsection{QTDE Framework}

Let $X$ be a dataset whose elements carry relational structure (for instance a
graph, or a network). We associate to $X$ an abstract simplicial
complex $K(X)$: a finite collection of simplices closed under taking faces,
where a $k$-simplex $\sigma=[v_0,\dots,v_k]$ records a $k$-th order interaction
among $k+1$ elements. Throughout, $K(X)$ is the clique (flag) complex of the
underlying graph, so that a $k$-simplex is a $(k{+}1)$-clique.

We propose two representations of $X$ built from~\Cref{eq:hodge-laplacian} and~\Cref{eq:evolution}: an \emph{implicit} (kernel) representation in which the
datum is mapped to a quantum state and compared through inner products, and an
\emph{explicit} (measurement) representation in which the datum is mapped to a
finite real feature vector. Both share the same evolution; they differ only in
how the evolved state is read out.

\subsection{Instantiations of QTDE}
\subsubsection{Implicit Representation: Topological Quantum Kernel}
\label{sec:implicit}

We define the feature map $K(X)\mapsto\mathcal H$ that sends the complex to the
quantum state obtained by evolving a fixed reference state $\ket{s}$ under the
topology-driven unitary \eqref{eq:evolution},
\begin{equation}
  \ket{\Psi_k(X)} \;=\; U_k(t)\,\ket{s}
                 \;=\; e^{\,i\,L_k t}\,\ket{s},
  \qquad \ket{s}\in\mathbb{C}^{\,n_k},
  \label{eq:state}
\end{equation}
where $\ket{s}$ is the uniform superposition state over the simplices in the complex;
\begin{equation}
\ket{s}
=
\frac{1}{\sqrt{n_k}}
\sum_{i=1}^{n_k}
|i\rangle.
\label{eq:uniform}
\end{equation}

the state therefore has
the \emph{same dimension as the Laplacian}. The map is a genuine feature map:
the (in general high-dimensional) vector $\ket{\Psi_k(X)}$ is never materialised
explicitly and is accessed only through inner products. For two datasets
$X_1,X_2$ we compare the induced states by the fidelity kernel
\begin{equation}
\small
\begin{aligned}
\mathcal K_k(X_1,X_2)
&= \bigl|\,\langle \Psi_k(X_1)\,|\,\Psi_k(X_2)\rangle\,\bigr|^2 \\
&= \bigl|\,\bra{s}\,e^{-iL_k(X_1)t}\,e^{\,iL_k(X_2)t}\,\ket{s}\,\bigr|^2 .
\end{aligned}
\label{eq:kernel}
\end{equation}
Writing $G_{12}=\langle\Psi_k(X_1)|\Psi_k(X_2)\rangle$ for the Gram matrix of
the states, the kernel \eqref{eq:kernel} equals the entrywise product
$G\circ\overline G$; by the Schur product theorem it is positive semidefinite,
so it is a valid kernel and can be used directly in a support vector machine,
\begin{equation}
  f(X)=\operatorname{sign}\!\Big(\textstyle\sum_i \alpha_i y_i\,\mathcal K_k(X_i,X)+b\Big).
\end{equation}
Comparing complexes through \eqref{eq:kernel}, rather than comparing the
matrices $L_k$ directly, lets the unitary dynamics expose spectral features of
the higher-dimensional topology while keeping the comparison invariant to the
unobservable global phase. For two complexes with different numbers of
$k$-simplices the states~\eqref{eq:state} are embedded in a common Hilbert space
by zero-extension before the inner product~\eqref{eq:kernel} is taken; the
evolution always occurs in $\mathbb{C}^{n_k}$.

\subsubsection{Explicit Representation: Measurement Features}
\label{sec:explicit}

The implicit representation requires the full state and a pairwise kernel. The
second representation instead summarises the same evolution by a small number of
measurement outcomes, yielding an explicit feature vector amenable to a linear model.
Starting from a fixed initial state $\ket{s}$ (\Cref{eq:uniform}) on the $k$-simplices, we evolve
\begin{equation}
  \ket{s_X(t)} \;=\; e^{-i\,L_k(X)\,t}\ket{s} ,
\end{equation}
and record the expectation values of a chosen set of observables
$\{M_j\}_{j=1}^{m}$ at times $\{t_j\}_{j=1}^{m}$,
\begin{equation}
  \phi(X) \;=\; \big(\,\langle s_X(t_j),\,M_j\,s_X(t_j)\rangle\,\big)_{j=1}^{m}
           \;\in\;\mathbb{R}^{m}.
  \label{eq:features}
\end{equation}
The data are then classified \emph{linearly} in this explicit feature space,
\begin{equation}
  f(X)=\operatorname{sign}\!\big(w\cdot\phi(X)+b\big),
  \label{eq:linear}
\end{equation}
with weight vector $w$ and bias $b$ learned from data. A canonical and
inexpensive choice of observable is the projector onto the initial state,
$M=\ket{s}\!\bra{s}$, whose expectation is the
\emph{survival (return) amplitude}
\begin{equation}
  a(t)\;=\;\bra{s}e^{\,iL_k t}\ket{s}
        \;=\;\sum_{\lambda} \big|\langle s|\phi_\lambda\rangle\big|^2\,e^{\,i\lambda t},
  \label{eq:survival}
\end{equation}
where $\{(\lambda,\phi_\lambda)\}$ is the eigensystem of $L_k$. Equation~\eqref{eq:survival} is the characteristic function of the spectral measure of
$L_k$ seen from $\ket{s}$: its time-stationary component is supported on the
harmonic subspace $\ker L_k$ and thus carries the topological ($\beta_k$)
content, while its oscillatory component encodes the remainder of the spectrum.
Sampling $a(t)$ at several times and stacking
$\phi(X)=\big(\Re a(t_j),\,\Im a(t_j),\,|a(t_j)|^2\big)_j$ gives a compact,
basis-free fingerprint of the dimension-$k$ topology that requires neither a
shared Hilbert space nor pairwise kernel evaluations.

\subsection{Generalized Spectral Encodings}
\label{sec:qsvt}
The exponential filter \eqref{eq:evolution} is one element of a broader family.
Since $e^{-iL_k t}$ is a function of $L_k$, one may replace it by an arbitrary
(trainable) polynomial applied to the Laplacian,
\begin{equation}
  p_\theta(L_k)=\sum_{r=0}^{R} a_r\,L_k^{\,r},
\end{equation}
realisable on quantum hardware through quantum singular value transformation,
and use $p_\theta(L_k)\ket{s}$ in either representation. The exponential map is
recovered for $a_r=(-it)^r/r!$. Tunable coefficients let the filter emphasize
the spectral ranges of the combinatorial Laplacian that best separate the classes, at
the cost of additional parameters. The expansion
$e^{-ip_{\theta}(L_k) t}\ket{s}=\sum_{r\ge0}\tfrac{(-it)^r}{r!}p_\theta(L_k)^{\,r}\ket{s}$ makes explicit that
both representations implicitly probe high powers of a large, higher-dimensional
operator.

\section{Testing and Benchmarking}\label{sec:methodology}
\subsection{Implementation and Reproducibility}

All simulations were implemented in Python. Graphs and their clique (flag)
complexes were constructed with \textsc{NetworkX}, and the combinatorial
Laplacians $L_k$ were assembled as sparse matrices from the boundary operators
$B_k, B_{k+1}$ using \textsc{SciPy}. The action of the topology-driven unitary
$e^{-i L_k t}$ on the initial state was evaluated by a Krylov-subspace
matrix-exponential routine (\texttt{scipy.sparse.linalg.expm\_multiply}), which
exploits the sparsity of $L_k$ and returns the evolved state at all sampled
times $t_1, \dots, t_T$ in a single pass; this is the classical surrogate for
the Hamiltonian-simulation step discussed in Sec.~VI. Classification was
performed with the support vector machines of \textsc{scikit-learn}: the
fidelity and matrix-difference kernels were passed as precomputed Gram matrices,
while the explicit survival features were classified with a linear SVM. Figures
were produced with \textsc{Matplotlib}.

To keep the reported numbers reproducible, we fix all sources of randomness and
hold the learning pipeline constant across representations. The dataset is
generated from a fixed seed; the SVM regularization is fixed at $C=1$ for every
method; the Gaussian-kernel bandwidth of the classical baseline is set by the
median heuristic rather than tuned; and the survival features are standardized
inside the cross-validation pipeline (using training-fold statistics only) so
that no information leaks from validation folds. Model assessment uses stratified
five-fold cross-validation with a fixed fold assignment shared by all methods,
and we report the mean and standard deviation of the fold accuracies. Each simplicial dimension $k$ is evaluated as a separate classification problem, with no averaging across dimensions, so that the discriminative contribution of each topological scale can be read off independently.

The complete source code needed to reproduce the datasets, encodings, baselines,
and figures will be made publicly available as open source upon publication.

\subsection{Benchmark Datasets}

To evaluate the proposed QTDE framework, we consider a binary classification task on random topological spaces: clique complexes based on Erd\H{o}s--R\'enyi random graphs with different densities. Two classes are generated according to different edge probabilities,
$$
G(n,p_0), \qquad G(n,p_1),
$$
with a common vertex set of size $n$. The classification task therefore isolates differences in graph density while preserving the vertex identities across all samples. For each graph, the associated clique complex (flag complex) is constructed. A clique of size $(k+1)$ is interpreted as a $k$-simplex. Simplices are enumerated up to a prescribed maximum dimension $k_{\max}$. 
Our tests constitute an elementary benchmarking methodology, and serve as a proof of concept rather than an extensive and comprehensive benchmarking on domain-specific datasets, which we reserve for future works.

\subsection{Construction of QTDE Representations}

The Hilbert space associated with a graph at dimension $k$ is
$\mathcal H_k \cong \mathbb C^{n_k},$ where $n_k = |S_k|$ is the number of $k$-simplices. Each basis state corresponds to a single simplex. We use a uniform superposition over all $k$-simplices, as a starting state \Cref{eq:uniform}. This initialization ensures that all simplices contribute equally to the dynamics and yields a global spectral probe of the Laplacian. For numerical stability while preserving relative scale information, all Laplacians at a fixed dimension are normalized using a single dataset-wide constant,
$
\widetilde L_k
=
\frac{L_k}{c_k},
$
where
$
c_k
=
\max_{G}
\left(
\max \operatorname{diag}(L_k(G))
\right).
$
The quantum state associated with graph $G$ is evolved according to
\begin{equation}
|\Psi_k(t)\rangle
=
e^{-i\widetilde L_k t}
\ket{s} ,
\label{eq:timedependentqstate}
\end{equation}
for a discrete set of evolution times
$t_1,\ldots,t_T.$
The matrix exponential action is evaluated using sparse Krylov-based propagation~\cite{Park_Light_1986Krylov}.

\subsection{Baselines}

We know that encoding the topological data into quantum states is an interesting research direction on its own. Direct comparison of the Laplacian matrices is a natural baseline, but it might be hard to encode that data on quantum computers and manipulating large matrices is something that quantum computers might struggle with, whereas, on the other side manipulating quantum states and implementing quantum evolution are natural quantum computing operations.
To assess whether the topology-driven quantum evolution serves as a more informative representation than the combinatorial
Laplacian, we compare QTDE against a classical baseline that operates on the
\emph{same} object, $L_k$, but compares data by direct matrix comparison rather
than through quantum dynamics.

For a fixed dimension $k$, each datum $X$ is represented by its combinatorial
Laplacian $L_k(X)$. Because all graphs in a given benchmark share a common
vertex set, their $k$-simplices are identified across samples, and every
$L_k(X)$ is embedded in a common $N \times N$ simplex basis, with $N$ the
number of distinct $k$-simplices. The similarity between two data is
then measured by a Gaussian (radial basis function) kernel on the Frobenius
distance between their Laplacians,
\begin{equation}
  K_k^{\mathrm{F}}(X_i, X_j)
  = \exp\!\big(-\gamma\, \lVert L_k(X_i) - L_k(X_j) \rVert_F^2 \big),
\end{equation}
where $\lVert \cdot \rVert_F$ is the Frobenius norm and the bandwidth is fixed
by the median heuristic, $\gamma = 1 / \mathrm{median}_{i<j} \lVert L_k(X_i) -
L_k(X_j) \rVert_F^2$. Since the Laplacians live in the Euclidean space
$(\mathbb{R}^{N \times N}, \langle \cdot, \cdot \rangle_F)$, $K_k^{\mathrm{F}}$
is a standard Gaussian kernel and is therefore positive semidefinite; it can be
used directly in the same support vector machine as the quantum kernels, so that only the representation, and not the classifier, differs
across methods.

This is a natural choice of baseline for three reasons. First, it is the most
direct classical realization of the comparison that QTDE is proposed to replace:
the implicit representation was introduced precisely as an alternative to
comparing the matrices $L_k$ directly (Sec.~III\,B), and $K_k^{\mathrm{F}}$ is
exactly that direct comparison. The baseline therefore isolates the contribution
of the quantum evolution $e^{-iL_k t}$, since both approaches take identical
inputs (the combinatorial Laplacian at the same dimension $k$) and differ only
in whether the topology is compared as a static operator or through the state it
generates. Second, the Frobenius metric is the canonical, basis-consistent
distance on the space of Laplacians, and combined with the Gaussian kernel it
yields a valid positive-semidefinite kernel usable in the identical SVM
pipeline; this holds the learning algorithm fixed and attributes any performance
gap to the representation alone. Third, the baseline is simple and
nearly free of hyperparameters, making it a
conservative reference: an improvement over $K_k^{\mathrm{F}}$ cannot be
ascribed to additional feature engineering or tuning on the classical side.

\subsection{Evaluation Protocol}

All experiments that test the accuracy of classification employ stratified five-fold cross-validation (CV). For kernel-based methods, the complete kernel matrix is computed once. For explicit features, standardization is performed inside the training pipeline to prevent information leakage. The similarity matrix for explicit feature vectors, as in \Cref{eq:survival}, is computed based on the standard measure \textit{cosine similarity}, which evaluates the correlation between the two vectors, and equals 1 for identical vectors pointing in the same direction, $-1$ for opposite direction, and 0 for two completely independent vectors.

Classification performance is reported as the mean and standard deviation of the cross-validation accuracy. Results are presented separately for each dimension $k$, enabling a detailed analysis of how different topological scales contribute to discrimination between the topologically different regimes.

\section{Results}\label{sec:results}

\subsection{QTDE vs. Classical Laplacian Comparisons}

\begin{figure*}[t!]
    \centering
    \includegraphics[width=1\linewidth]{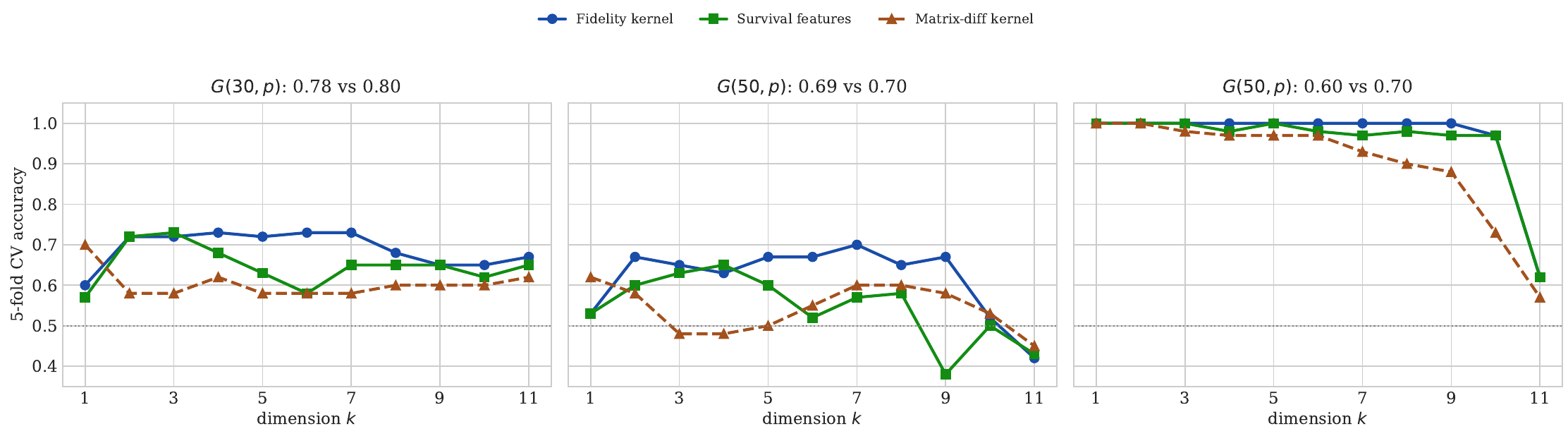}
    \caption{\justifying Classification accuracy based on QTDE (using a support vector machine) as a function of simplicial dimension $k$ for binary classification tasks on clique complexes based on Erd\H{o}s--R\'enyi random graph ensembles. The three panels correspond to increasingly challenging discrimination tasks between topological-space classes generated based on random graphs with different edge probabilities. Results are shown for the proposed fidelity kernel representation (blue) and survival features (green), together with a classical baseline based on direct comparison of combinatorial Laplacians through a matrix-difference kernel (brown). Classification performance is reported as the mean 5-fold cross-validation accuracy.
}
    \label{fig:qtde_accuracy}
\end{figure*}

The three panels of~\Cref{fig:qtde_accuracy} form a density-gap ladder that fixes
the intrinsic difficulty of the task. The widest gap, $G(50,0.60)$ vs.\
$G(50,0.70)$, is almost perfectly separable, all three representations sit near
unit accuracy over most of the dimension range, whereas the more similar
ensembles $G(30,0.78)$ vs.\ $G(30,0.80)$ and, most acutely, $G(50,0.69)$ vs.\
$G(50,0.70)$ compress every method into the $0.5$--$0.7$ band. Separability tracks the difference in edge probabilities $p$, and the two small-gap panels represent a genuinely hard discrimination
problem.

The comparison between the quantum and classical representations is most
informative on the separable panel, and specifically at large $k$. At low and
intermediate dimensions all three curves are saturated and indistinguishable, so
that regime carries little discriminative information about the representations
themselves. As $k$ increases, however, the matrix-difference baseline falls away
while the fidelity kernel and survival features retain high accuracy for several
additional dimensions before eventually declining. 
The sudden drop in the top dimensions, especially pronounced for smaller values of $p$, should be attributed to the fact that for sufficiently high $k$ with relatively small vertex set $n$ the $L_k$ combinatorial Laplacians of the \textit{clique complexes} do not contain many $k$-dimensional simplices in the underlying graph and are close to being zero matrices (See \Cref{sec: clique}).

On the two hard panels the picture is more uniform: the fidelity, survival, and
matrix-difference curves interleave within the $0.5$--$0.7$ band, and the
separations between them are comparable to the cross-validation spread of the
individual points. The quantum readouts are generally at or above the classical
baseline across these dimensions, with the survival features the steadier of the
two. The fidelity kernel exhibits larger dimension-to-dimension fluctuations and
occasional dips toward chance but we do not read any single dimension as
decisive on these ensembles. Across all three tasks the accuracy is clearly
non-monotonic in $k$: the most discriminative simplicial scale is
task-dependent rather than simply the largest available dimension, so the value
of the higher-order Laplacians is realised only when the relevant topological
structure is present. 


\subsection{Implicit vs Explicit Representations}
\label{sec:impvsexp}
\begin{figure*}[t]
    \centering

        \begin{subfigure}[t]{0.48\textwidth}
        \centering
        \includegraphics[width=\linewidth]{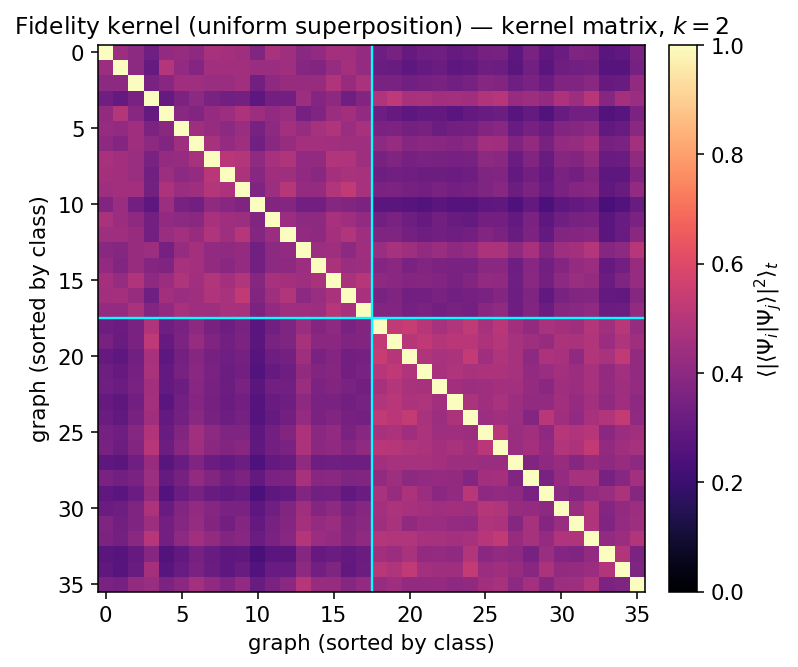}
        \caption{}
        \label{fig:kernel_fidelity_k2}
    \end{subfigure}
    \hfill
    \begin{subfigure}[t]{0.48\textwidth}
        \centering
        \includegraphics[width=\linewidth]{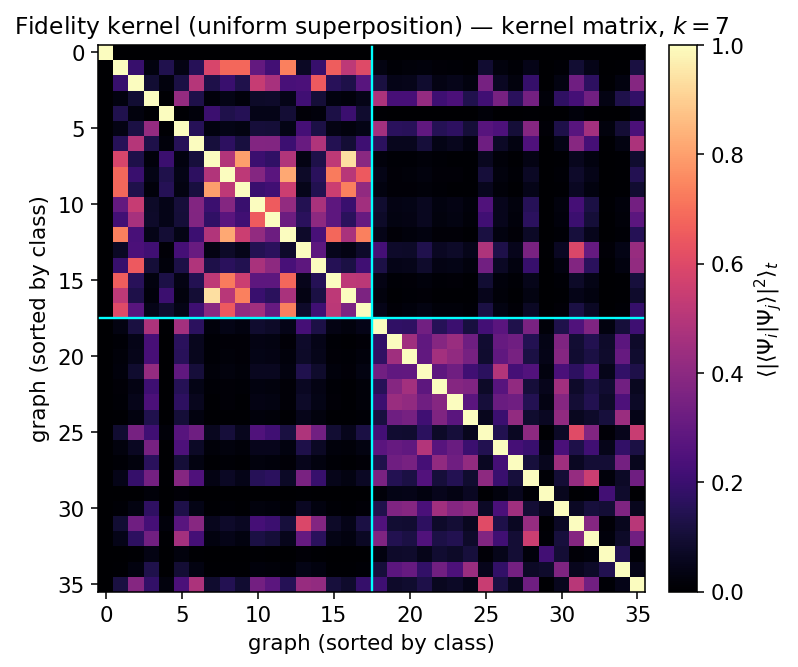}
        \caption{}
        \label{fig:kernel_fidelity_k7}
    \end{subfigure}

    \vspace{0.5cm}

 \begin{subfigure}[t]{0.48\textwidth}
        \centering
        \includegraphics[width=\linewidth]{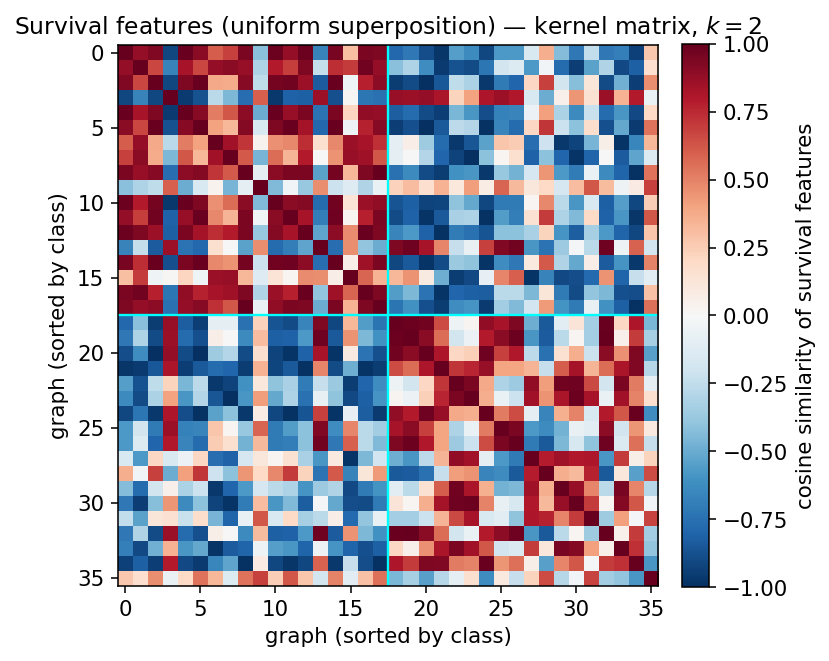}
        \caption{}
        \label{fig:kernel_survival_k2}
    \end{subfigure}
    \hfill
    \begin{subfigure}[t]{0.48\textwidth}
        \centering
        \includegraphics[width=\linewidth]{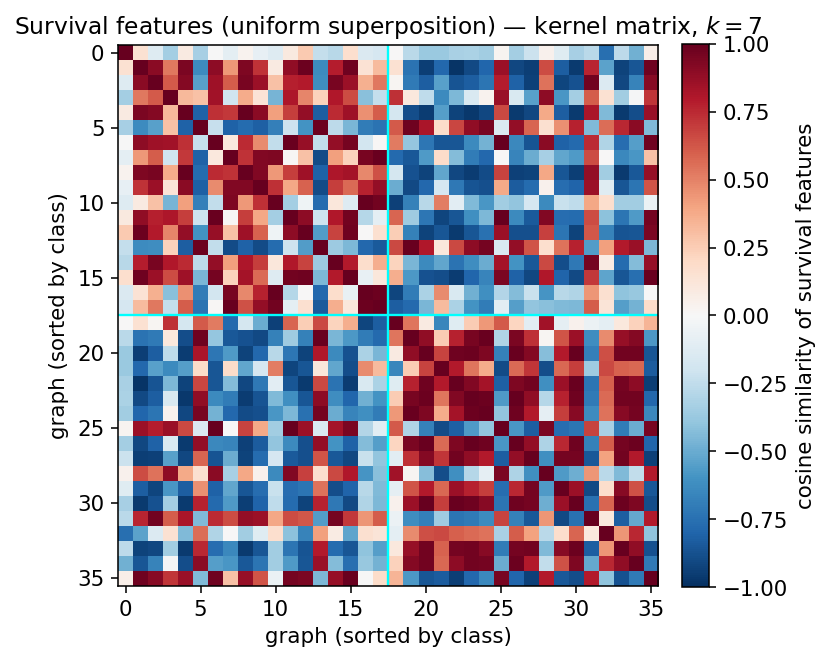}
        \caption{}
        \label{fig:kernel_survival_k7}
    \end{subfigure}

\caption{\justifying Comparison of implicit and explicit representations produced by QTDE. (a) and (b) show the fidelity kernel matrices computed from the evolved quantum states for simplicial dimensions $k=2$ and $k=7$, respectively. (c) and (d) show the corresponding similarity matrices obtained from the explicit survival-feature representations using cosine similarity. Graphs are ordered by class, cyan lines indicate the class boundary separating the two graph ensembles.
}
    \label{fig:kernel_comparison}
\end{figure*}

\Cref{fig:kernel_comparison} compares the similarity structures induced by the two representations. Although the implicit and explicit approaches access the topology-driven quantum evolution in fundamentally different ways, both produce similarity matrices that clearly separate the two graph classes. In particular, the block-diagonal structure visible in all four panels indicates that graphs belonging to the same class are assigned higher similarity than graphs from different classes. This observation demonstrates that the two representations preserve essentially the same global organization of the dataset despite relying on different information extracted from the quantum evolution.

The effect of the simplicial dimension is particularly apparent when comparing the left and right columns. At $k=2$, the similarity matrices exhibit relatively homogeneous within-class structure, whereas at $k=7$ substantially richer patterns emerge. The fidelity kernel produces a more localized similarity structure, with stronger contrast between similar and dissimilar graphs, while the explicit survival features yield a smoother representation that nevertheless preserves the same class separation. These observations suggest that higher-order simplicial interactions encode additional discriminative information and that the explicit measurement-based representation retains much of the information contained in the full quantum state kernel while providing a significantly lower-dimensional description.


\subsection{QSVT Spectral Filters}
\label{sec:qsvt-results}

\begin{figure*}[ht]
    \centering
    \includegraphics[width=\linewidth]{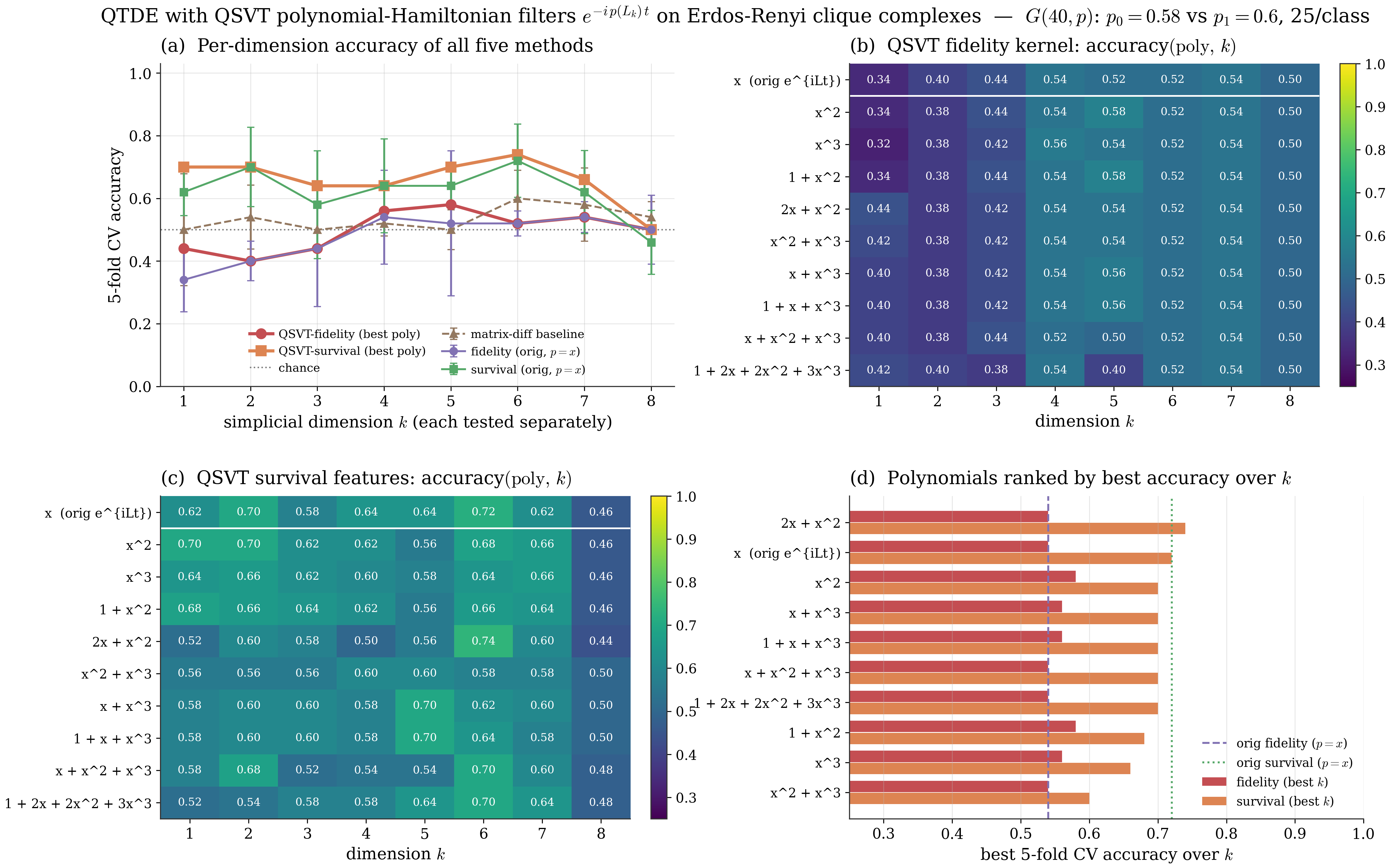}
    \caption{\justifying QSVT with degree-three polynomial spectral filters $p_\theta(L_k)$ on
    Erd\H{o}s--R\'enyi clique complexes ($G(40,p)$: $p_0=0.58$ vs.\ $p_1=0.60$,
    25 graphs per class). (a) Per-dimension accuracy for the
    best-performing filter at each $k$, for the fidelity kernel and survival
    features, against the original exponential encoding and the
    matrix-difference baseline. Figures (b, c) accuracy as a function of filter and
    dimension $k$ for the fidelity kernel and survival features, respectively.
    (d) Filters ranked by best accuracy attained over all $k$. Error bars in
    (a) denote the standard deviation over five-fold cross-validation.}
    \label{fig:qsvt(2)}
\end{figure*}

Figure~\ref{fig:qsvt(2)} tests whether replacing the exponential evolution
$e^{-iL_kt}$ with a trainable polynomial filter $p_\theta(L_k)$
(see \Cref{sec:qsvt}) yields a more discriminative representation
than the fixed Hamiltonian encoding, on the harder $G(40,p)$ benchmark
($p_0=0.58$ vs.\ $p_1=0.60$, 25 graphs per class). For each simplicial
dimension $k=1,\dots,8$ we scan a fixed library of ten
polynomials $p(x)$ (Table in panel (b)/(c) row labels), implement each as
a QSVT phase sequence acting on $L_k$, and evaluate both readouts (the fidelity kernel $\kappa_k$ and the survival-feature vector
$\phi(X)$) against the original exponential encoding and the
matrix-difference baseline $K^F_k$.

\paragraph{Per-dimension comparison of best filters (panel a).}
Selecting, at each $k$, the best-performing polynomial from the library
(\emph{QSVT-fidelity, best poly} and \emph{QSVT-survival, best poly})
produces a curve that sits consistently above both the original
exponential fidelity/survival kernels and the matrix-difference baseline
for nearly the entire range $k=1,\dots,8$, with the largest separation
in the intermediate regime $k\approx4$--$6$. 

\paragraph{Polynomial-by-dimension structure (panels b, c).}
The heatmaps in panels (b) and (c) show accuracy as a joint function of
polynomial choice and $k$, for the fidelity kernel and survival features
respectively. Two features stand out. First, accuracy is not uniform
across the polynomial library at fixed $k$: at the dimensions where
QSVT provides its largest gain, purely even- or high-degree filters
(e.g.\ $x^2$, $x^3$) noticeably outperform the linear filter $x$ (the
original exponential encoding's generator), indicating that
compressing the low-lying part of the spectrum is what drives the improvement rather than an arbitrary reparametrisation of the evolution time. Second, the
two panels are qualitatively very similar to one another: the
polynomials that are strong for the fidelity kernel at a given $k$ are,
with few exceptions, also strong for the survival features at that same
$k$. This mirrors the observation in \Cref{sec:impvsexp}
(\Cref{fig:qtde_accuracy}) that the implicit and explicit readouts extract
largely overlapping information from the evolved state; the QSVT
generalisation does not appear to break that equivalence, which suggests
the benefit of spectral filtering is a property of the encoding
$p_\theta(L_k)\,|s\rangle$ itself, not an artifact of how the resulting
state is subsequently measured.

\paragraph{Aggregate polynomial ranking (panel d).}
Ranking each polynomial by its best accuracy attained over all $k$
(panel D) shows that no single filter dominates uniformly: the top few
entries are separated by a small margin, and several of them
(including $2x+x^2$ and the plain quadratic $x^2$) outperform the
original linear filter $x$, for both the fidelity-kernel and
survival-feature readouts. This is consistent with the claim that
\emph{some} member of a modestly sized degree-three filter family is
generally at least as good as the exponential evolution, even though
identifying which member is best a priori is not obvious and appears to
depend on $k$ and on the density gap of the underlying graph.

\begin{figure}[t]
  \centering
  \includegraphics[width=\linewidth]{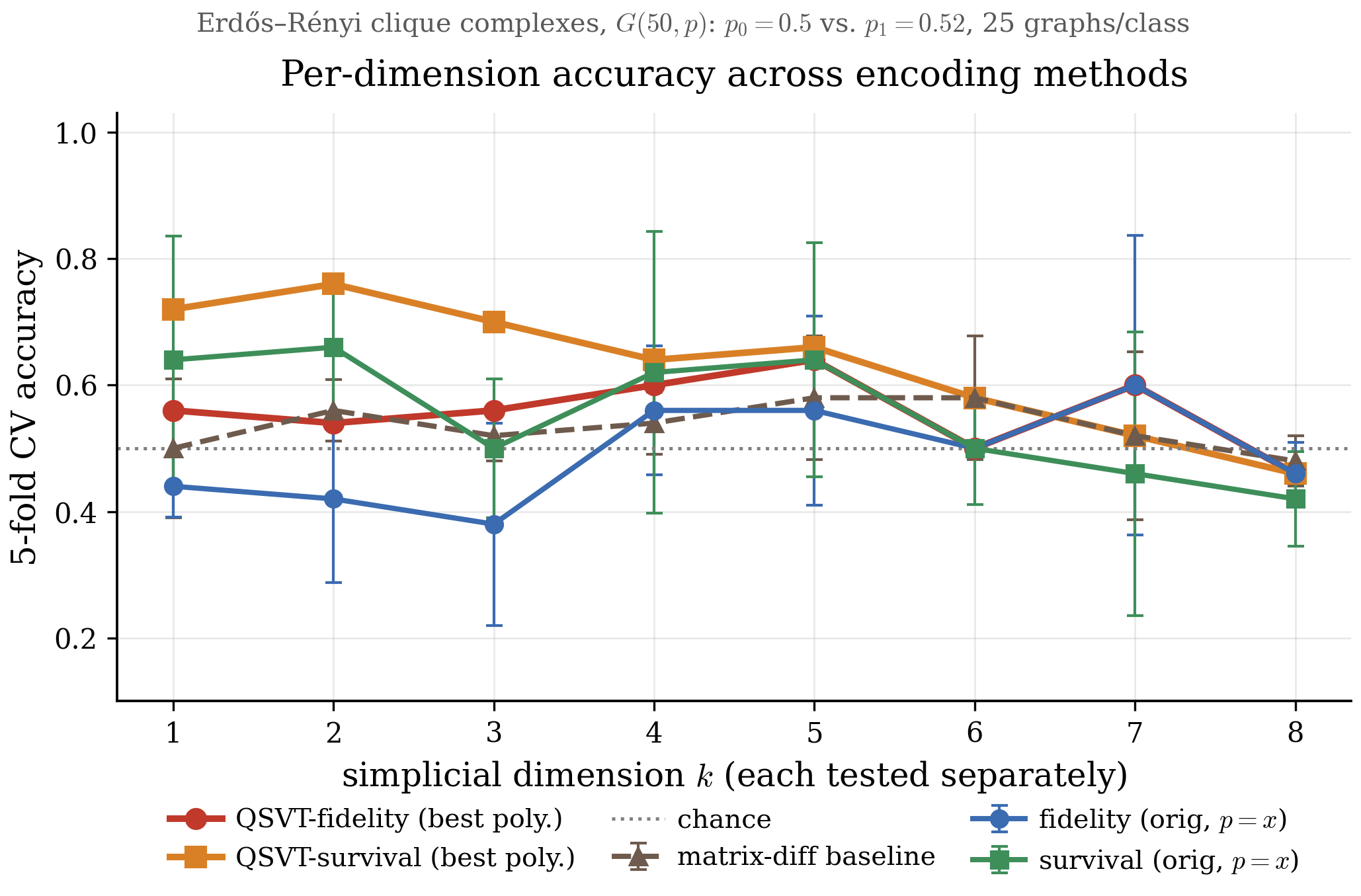}
  \caption{\justifying Per-dimension classification accuracy of the five QTDE encodings. Five-fold cross-validated accuracy for discriminating
    Erd\H{o}s--R\'enyi clique complexes $G(50,0.5)$ from $G(50,0.52)$
    ($25$ graphs per class), with each simplicial dimension $k$ classified
    independently. We compare the classical matrix-difference baseline, the original exponential
    evolution $e^{-iL_k t}$ read out via the fidelity kernel and via
    survival features (both $p(x)=x$), and the corresponding QSVT
    polynomial-Hamiltonian filters $e^{-i\,p(L_k)\,t}$. For the two QSVT
    curves, the value at each $k$ is the best accuracy over a family of ten
    polynomial filters. Markers denote the mean over folds and error
    bars the standard deviation; the dotted line marks chance ($0.5$).
    }
  \label{fig:acc-vs-dim}
\end{figure}

\subsection{Other polynomials}
In \Cref{fig:acc-vs-dim} we report the result of an extended QSVT sweep over filters drawn from four
polynomial families: (i) the \textit{Chebyshev polynomials} $T_1(x)=x$,
$T_2(x)=2x^2-1$, $T_3(x)=4x^3-3x$, $T_4(x)=8x^4-8x^2+1$ and
$T_5(x)=16x^5-20x^3+5x$; (ii) the \textit{Hermite polynomial} $H_3(x)=8x^3-12x$; (iii) the \textit{cyclotomic
polynomials} $\Phi_5(x)=x^4+x^3+x^2+x+1$ and $\Phi_8(x)=x^4+1$; and (iv) the \textit{Fibonacci polynomials} $F_6(x)=x^5+4x^3+3x$ and
$F_{11}(x)=x^{10}+9x^8+28x^6+35x^4+15x^2+1$. The identity $p(x)=x$ reproduces the original exponential evolution
$e^{-iL_k t}$ and serves as the baseline.

Two trends survive the noise. First, the discriminative
signal is concentrated at low simplicial dimension. The survival-based
readouts peak at $k=1$--$2$ reaching a nominal $\approx 0.76$ for the
best QSVT polynomial and $\approx 0.66$ for the original $p(x)=x$
evolution and then decay monotonically towards, and eventually below,
chance by $k=6$--$8$, where high-dimensional cliques become sparse and
dominated by sampling noise. Second, the survival features are the only
encodings that rise clearly above chance, and only in this low-$k$ regime;
the fidelity kernel and the classical matrix-difference baseline remain
close to chance across all $k$.

The QSVT polynomial-Hamiltonian filters yield only marginal gains over the
original exponential evolution. For the fidelity kernel every polynomial clusters at the $p(x)=x$
reference ($\approx 0.60$), and for the survival features the best filters
exceed the $p(x)=x$ baseline by only
$0.05$--$0.10$. These QSVT scores are, moreover, the \emph{maximum}
accuracy over a family of polynomials at each $k$. This
selection bias inflates the reported values, so the residual advantages
fall within the cross-validation spread. We therefore read the results as
consistent with the previous conclusion: the evidence gathered on this task suggests the
polynomial-Hamiltonian generalisation offers marginal improvement that is hard to predict a priori and can sometimes lead to deterioration of results. Results for accuracy improvements for specific polynomials are reported in Appendix \ref{sec:specificpoly} in \Cref{fig:Specific_Poly}.

\subsection{Discussion}

The proposed framework differs from conventional quantum data encodings by constructing quantum states from the topology of the data rather than from its raw coordinates. The combinatorial Laplacian captures higher-order interactions among simplices, and the resulting quantum evolution transforms these interactions into interference patterns that depend on the spectrum of $L_k$. Consequently, two datasets with similar higher-order topology produce similar quantum states even when their individual graph representations differ substantially. Unlike vector-based encodings, QTDE is invariant to vertex permutations once the simplicial complex is fixed, making the representation naturally suited to relational data.

The exponential evolution $e^{-iL_kt}$ represents only one particular spectral transformation of the combinatorial Laplacian. Replacing the exponential with polynomial spectral filters implemented through QSVT gives flexibility to amplify or suppress different regions of the Laplacian spectrum.
The empirical results indicate that several polynomial filters consistently outperform the original exponential evolution. Although the absolute improvements are modest, they show that learning or optimizing the spectral response can produce more discriminative quantum representations than fixed Hamiltonian evolution.

Our experiments further show that increasing the simplicial dimension $k$ does not monotonically improve classification performance (\Cref{fig:qtde_accuracy}).
Instead, different datasets appear to possess informative topology at different dimensions, and the optimal choice of $k$ depends on the interaction structure relevant to the learning task. Lower-dimensional Laplacians primarily capture local connectivity, whereas higher-dimensional Laplacians encode interactions among larger collections of simplices. When such higher-order structures are present, larger values of $k$ provide additional discriminative information; otherwise, they may introduce unnecessary complexity without improving prediction. Despite relying on different readout mechanisms, the implicit fidelity kernel and explicit survival features produce remarkably similar similarity structures and comparable classification performance. This suggests that much of the information contained in the evolved quantum state can be recovered from a relatively small collection of measurement-derived observables, offering a favorable trade-off between representational power and computational cost.

Several limitations remain. First, the size of the combinatorial Laplacians grows rapidly with the simplicial dimension, limiting scalability for dense complexes. Second, the current work considers fixed polynomial filters rather than filters optimized directly for downstream learning objectives. Finally, the benchmarks considered here are relatively small, and evaluating QTDE on larger datasets with clear practical meaning will be an important direction for future work. A natural direction is to learn spectral filters directly from data, combining QTDE with variational optimization or gradient-based QSVT synthesis. More broadly, the framework naturally extends to persistent simplicial complexes and dynamic topological structures, suggesting a general family of topology-aware quantum representation learning methods.

\section{Computational considerations}\label{sec:efficiency}
The computational requirements of QTDE arise from three sources: the size of the Hilbert-space representation, the implementation of topology-driven quantum evolution, and the cost of constructing sparse access to the combinatorial Laplacian.

\paragraph{Representation Size.}

The Laplacian matrices in dimension $k$ are $n \choose k+1 $$\times$$n \choose k+1 $ matrices, thus similarly the matrix $U=e^{-ip_{\theta}(L_k)t}$ is of the same size. The vectors $\ket{\Psi(\mathcal{T})}$ have to match the column size of $U$, which is polynomial in $n$, for any constant dimension $k=1,2,3...$ as well as $k=n-1, n-2, n-3...$. More efficient encodings with respect to the number of qubits are possible if one considers reduced Laplacians, but to enact the quantum evolution as described in this work the qubit requirements scale linearly with the number of $k$-simplices.

\paragraph{Quantum Implementation.}

In terms of time and circuit complexity, the efficiency of the quantum evolution step reduces to the efficiency of implementing the unitary evolution generated by $L_k$, or, in the QSVT variant, applying a polynomial spectral filter derived from $p_\theta$. Standard Hamiltonian simulation methods such as Suzuki--Trotter approximation~\cite{lloyd1996universal} can be employed. However, in our setting, methods based on Linear Combination of Unitaries~\cite{LCU}, or block-encoding in general, and related sparse-Hamiltonian simulation techniques~\cite{LCUberry2015hamiltonian} are more natural because they can exploit the sparsity of the combinatorial Laplacian $L_k$.  We note that there also exists efficient (depth logarithmic in the number of vertices) block encodings of the simplicial Dirac operator, which squares to the combinatorial Laplacian~\cite{kerenidis2022quantummachinelearningsubspace}. An efficient implementation of discrete-time quantum walks on simplicial complexes was recently introduced in Ref.~\cite{Hayakawa2026quantumwalks}.

In the block-encoding approach, the first step is not to construct the full evolution operator $e^{-iL_kt}$ directly. Instead, one constructs a unitary $U_L$ that encodes a scaled version of the Laplacian as one of its blocks, for example
$(\langle 0|\otimes I)U_L(|0\rangle\otimes I)=L_k/\alpha$,
where $\alpha$ is a normalization factor chosen so that $L_k/\alpha$ can be block-encoded. This block-encoding is independent of the evolution time $t$. To implement a particular evolution $e^{-iL_kt}$, one then uses Hamiltonian simulation methods based on the same $U_L$ and $U_L^\dagger$, with additional phase rotations chosen as a function of $\alpha t$ and the target precision. Thus, different evolution times require different phase choices and generally different numbers of calls to $U_L$, but they do not require reconstructing the block-encoding of $L_k$. The same logic applies to QSVT: a polynomial spectral filter is implemented by choosing a QSVT phase sequence for the corresponding polynomial on the scaled operator $L_k/\alpha$, while reusing the same block-encoding $U_L$. A degree-$R$ QSVT filter requires $O(R)$ calls to $U_L$ and $U_L^\dagger$. This follows from the standard pipeline for Hamiltonian simulation and QSVT~\cite{Gily_n_2019}.

\paragraph{Resource Requirements.}

To make the scaling explicit, let $n_k=|S_k|$ and let $s_k$ be the maximum number of nonzero entries in any row of $L_k$. One standard sparse-Hamiltonian implementation uses access to these nonzero entries: given a row and an index $\ell \leq s_k$, one can query the position and value of the $\ell$-th nonzero entry. Using this access procedure as a subroutine, the post-access cost of the quantum evolution can scale logarithmically in the Hilbert-space dimension $n_k$ and polynomially in $s_k$, the scaled evolution time $\alpha t$, and the precision parameters~\cite{LCUberry2015hamiltonian,Low_2019}. The total encoding cost therefore also includes the cost of constructing or providing this sparse access. If $L_k$ is built explicitly after the relevant simplices have been enumerated, this sparse preprocessing scales as $O(s_kn_k)$. Thus, for fixed $k$, fixed QSVT degree $R$, and a fixed number of measured features, constructing or providing access to $L_k$ is a leading per-instance cost; once this access is available, the same encoded operator can be reused for subsequent evolutions and QSVT filters. Nevertheless, for our classical simulation we use the Krylov-subspace method~\cite{Park_Light_1986Krylov} as noted in~\Cref{sec:methodology}.

\section{Summary and Outlook}\label{sec:conclusion}

In this work we introduced \emph{quantum topological data encoding}, QTDE, a general framework for representing topological data in quantum systems through topology-driven quantum evolution. Building upon and extending previous approaches for quantum representations of topological objects, our framework enables the encoding of higher-dimensional simplicial structures into quantum states by exploiting the combinatorial Laplacian associated with a simplicial complex. The framework naturally gives rise to two complementary learning paradigms: an implicit representation based on quantum kernels derived from state fidelities, and an explicit representation based on measurement-derived survival features. We further generalized this construction by introducing polynomial spectral transformations implemented through QSVT. In this way, QTDE provides a mechanism for transforming rich topological information into representations that can be processed by quantum machine learning algorithms.

We evaluated the proposed framework on several classification tasks involving topological data and compared its performance against a classical baseline based on direct comparisons of combinatorial Laplacians. Across the considered datasets, the quantum representations generated by QTDE consistently yielded improved classification performance, indicating that quantum evolutions driven by topological operators can capture structural features that are not readily accessible through conventional similarity measures. These results suggest that topology-driven quantum encodings constitute a promising direction for integrating ideas from topological data analysis and quantum machine learning.

At the same time, several limitations of the present study should be acknowledged. First, our empirical evaluation was restricted to a finite collection of benchmark datasets and relatively small problem instances that are compatible with classical simulation of quantum systems. Consequently, the extent to which the observed advantages persist for larger and more complex datasets remains an open question. Second, although the proposed framework demonstrates improved predictive performance relative to the considered classical baseline, the precise origin of this advantage is not yet fully understood. In particular, we do not provide formal guarantees regarding quantum advantage, expressive power, or generalisation performance. Finally, the computational cost associated with constructing simplicial complexes and combinatorial Laplacians may become substantial for large-scale datasets, motivating the development of more efficient preprocessing and encoding procedures.

These limitations point to several directions for future research. An important next step is the implementation and validation of QTDE on near-term quantum hardware in order to assess its robustness in the presence of realistic noise and hardware constraints~\cite{akhalwaya2024topological}. From a theoretical perspective, it will be valuable to characterise the expressive capabilities of topology-driven quantum encodings, establish complexity-theoretic guarantees, and identify settings in which genuine quantum advantages may arise~\cite{berry2024analyzingprospects}. Further work should also investigate alternative topological operators such as the discrete Dirac operator~\cite{Calmon_2023,wee2023persistent}, persistent and multi-scale topological constructions~\cite{pun2022persistent}, and adaptive or trainable topology-driven evolutions. Finally, extending QTDE to applications involving dynamic, temporal, or heterogeneous topological data, such as molecular systems, biological networks, and complex social interactions, may reveal new opportunities for quantum-enhanced learning~\cite{carlsson2020topological}.

Overall, our results demonstrate that topological structure can serve as the foundation of a quantum data encoding framework. We believe that quantum topological data encoding provides a versatile approach for exploiting higher-order structure in quantum machine learning and opens new avenues for research at the intersection of quantum computing and topological data analysis.

\section*{Acknowledgments}
L.P. is supported by the National Research Foundation, Singapore through the National Quantum Office, hosted in A*STAR, under its Centre for Quantum Technologies Funding Initiative (S24Q2d0009). L.P. is also partially supported by A*STAR under its Young Investigator Research Grant (YIRG) M25N8c0131. D.L. acknowledges support from the Ministry of Education of Singapore under its  SUTD Kickstarter Initiative (Grant No. SKI 20210501). We thank Vedran Dunjko for his valuable feedback and comments on an earlier draft of the manuscript. We are also grateful to Caesnan M.G. Leditto and Mahtab Yaghubi Rad for insightful discussions. We especially thank Caesnan M.G. Leditto for drawing our attention to important aspects of the QSVT extension technique. Finally, we thank the Centre for Quantum Technologies in Singapore for hosting Hackamonth 2026, where this project was initiated.

\vspace{0.4cm}
\emph{Code and Data Availability:} All numerical experiments are reproducible from the public repository at \url{https://github.com/DimitThanos/QTDE}.

\bibliography{bibliography}

@inproceedings{10.1145/2488608.2488720,
	title        = {Inverting well conditioned matrices in quantum logspace},
	author       = {Ta-Shma, Amnon},
	year         = {2013},
	booktitle    = {Proceedings of the Forty-Fifth Annual ACM Symposium on Theory of Computing},
	location     = {Palo Alto, California, USA},
	publisher    = {Association for Computing Machinery},
	address      = {New York, NY, USA},
	series       = {STOC '13},
	pages        = {881–890},
	doi          = {10.1145/2488608.2488720},
	isbn         = {9781450320290},
	url          = {https://doi.org/10.1145/2488608.2488720},
	numpages     = {10}
}

@inproceedings{aimeur2006machinelearning,
	title        = {Machine Learning in a Quantum World},
	author       = {A{\"i}meur, Esma and Brassard, Gilles and Gambs, S{\'e}bastien},
	year         = {2006},
	booktitle    = {Advances in Artificial Intelligence},
	publisher    = {Springer Berlin Heidelberg},
	address      = {Berlin, Heidelberg},
	pages        = {431--442},
	isbn         = {978-3-540-34630-2},
	editor       = {Lamontagne, Luc and Marchand, Mario}
}

@inproceedings{akhalwaya2024topological,
	title        = {Topological data analysis on noisy quantum computers},
	author       = {Akhalwaya, Ismail and Ubaru, Shashanka and Clarkson, Kenneth and Squillante, Mark and Jejjala, Vishnu and He, Yang-Hui and Naidoo, Kugendran and Kalantzis, Vasileios and Horesh, Lior},
	year         = {2024},
	booktitle    = {International Conference on Learning Representations},
	volume       = {2024},
	pages        = {34945--34978}
}

@article{beer2020training,
   title={Training deep quantum neural networks},
   volume={11},
   ISSN={2041-1723},
   url={http://dx.doi.org/10.1038/s41467-020-14454-2},
   DOI={10.1038/s41467-020-14454-2},
   number={1},
   journal={Nature Communications},
   publisher={Springer Science and Business Media LLC},
   author={Beer, Kerstin and Bondarenko, Dmytro and Farrelly, Terry and Osborne, Tobias J. and Salzmann, Robert and Scheiermann, Daniel and Wolf, Ramona},
   year={2020},
   month=Feb
}

@article{benioff1980computer,
	title        = {The computer as a physical system: A microscopic quantum mechanical Hamiltonian model of computers as represented by Turing machines},
	author       = {Benioff, Paul},
	year         = {1980},
	journal      = {Journal of Statistical Physics},
	publisher    = {Springer},
	volume       = {22},
	number       = {5},
	pages        = {563--591},
	doi          = {10.1007/BF01011339},
	url          = {https://doi.org/10.1007/BF01011339}
}

@article{berry2024analyzingprospects,
	title        = {Analyzing Prospects for Quantum Advantage in Topological Data Analysis},
	author       = {Berry, Dominic W. and Su, Yuan and Gyurik, Casper and King, Robbie and Basso, Joao and Barba, Alexander Del Toro and Rajput, Abhishek and Wiebe, Nathan and Dunjko, Vedran and Babbush, Ryan},
	year         = {2024},
	month        = {Feb},
	journal      = {PRX Quantum},
	publisher    = {American Physical Society},
	volume       = {5},
	pages        = {010319},
	doi          = {10.1103/PRXQuantum.5.010319},
	url          = {https://link.aps.org/doi/10.1103/PRXQuantum.5.010319},
	issue        = {1},
	numpages     = {45}
}

@book{bishop2006pattern,
	title        = {Pattern Recognition and Machine Learning},
	author       = {Bishop, Christopher M.},
	year         = {2006},
	publisher    = {Springer},
	address      = {New York, NY},
	isbn         = {978-0-387-31073-2},
	url          = {https://link.springer.com/book/9780387310732}
}

@article{Calmon_2023,
	title        = {Dirac signal processing of higher-order topological signals},
	author       = {Calmon, Lucille and Schaub, Michael T and Bianconi, Ginestra},
	year         = {2023},
	month        = {sep},
	journal      = {New Journal of Physics},
	publisher    = {IOP Publishing},
	volume       = {25},
	number       = {9},
	pages        = {093013},
	doi          = {10.1088/1367-2630/acf33c},
	url          = {https://doi.org/10.1088/1367-2630/acf33c}
}

@article{carlsson2009topology,
	title        = {Topology and Data},
	author       = {Gunnar Carlsson},
	year         = {2009},
	journal      = {Bulletin of the American Mathematical Society},
	volume       = {46},
	number       = {2},
	pages        = {255--308},
	doi          = {10.1090/S0273-0979-09-01249-X}
}

@article{carlsson2020topological,
	title        = {Topological methods for data modelling},
	author       = {Carlsson, Gunnar},
	year         = {2020},
	journal      = {Nature Reviews Physics},
	publisher    = {Nature Publishing Group UK London},
	volume       = {2},
	number       = {12},
	pages        = {697--708},
	url          = {https://doi.org/10.1038/s42254-020-00249-3}
}

@book{carlsson2021topological,
	title        = {Topological data analysis with applications},
	author       = {Carlsson, Gunnar and Vejdemo-Johansson, Mikael},
	year         = {2021},
	publisher    = {Cambridge University Press}
}

@article{caro2021encodingdependent,
	title        = {Encoding-dependent generalization bounds for parametrized quantum circuits},
	author       = {Caro, Matthias C. and Gil-Fuster, Elies and Meyer, Johannes Jakob and Eisert, Jens and Sweke, Ryan},
	year         = {2021},
	month        = nov,
	journal      = {{Quantum}},
	publisher    = {{Verein zur F{\"{o}}rderung des Open Access Publizierens in den Quantenwissenschaften}},
	volume       = {5},
	pages        = {582},
	doi          = {10.22331/q-2021-11-17-582},
	issn         = {2521-327X},
	url          = {https://doi.org/10.22331/q-2021-11-17-582}
}

@misc{chang2025primerquantummachinelearning,
	title        = {A Primer on Quantum Machine Learning},
	author       = {Su Yeon Chang and M. Cerezo},
	year         = {2025},
	url          = {https://arxiv.org/abs/2511.15969},
	eprint       = {2511.15969},
	archiveprefix = {arXiv},
	primaryclass = {quant-ph}
}

@article{chazal2021introduction,
	title        = {An introduction to topological data analysis: fundamental and practical aspects for data scientists},
	author       = {Chazal, Fr{\'e}d{\'e}ric and Michel, Bertrand},
	year         = {2021},
	journal      = {Frontiers in artificial intelligence},
	publisher    = {Frontiers Media SA},
	volume       = {4},
	pages        = {667963},
	doi          = {10.3389/frai.2021.667963},
	url          = {https://doi.org}
}

@article{dunjko2016quantumenhanced,
	title        = {Quantum-Enhanced Machine Learning},
	author       = {Dunjko, Vedran and Taylor, Jacob M. and Briegel, Hans J.},
	year         = {2016},
	month        = {Sep},
	journal      = {Phys. Rev. Lett.},
	publisher    = {American Physical Society},
	volume       = {117},
	pages        = {130501},
	doi          = {10.1103/PhysRevLett.117.130501},
	url          = {https://link.aps.org/doi/10.1103/PhysRevLett.117.130501},
	issue        = {13},
	numpages     = {6}
}

@article{dunjko2018machinelearning,
	title        = {Machine learning \& artificial intelligence in the quantum domain: a review of recent progress},
	author       = {Dunjko, Vedran and Briegel, Hans J},
	year         = {2018},
	month        = {jun},
	journal      = {Reports on Progress in Physics},
	publisher    = {IOP Publishing},
	volume       = {81},
	number       = {7},
	pages        = {074001},
	doi          = {10.1088/1361-6633/aab406},
	url          = {https://doi.org/10.1088/1361-6633/aab406}
}

@book{edelsbrunner2010computational,
	title        = {Computational Topology: An Introduction},
	author       = {Edelsbrunner, H. and Harer, J.},
	year         = {2010},
	publisher    = {American Mathematical Society},
	series       = {Applied Mathematics},
	isbn         = {9780821849255},
	url          = {https://books.google.com.sg/books?id=MDXa6gFRZuIC},
	lccn         = {2009028121}
}

@article{feynman1982simulating,
	title        = {Simulating physics with computers},
	author       = {Feynman, Richard P},
	year         = {1982},
	journal      = {International Journal of Theoretical Physics},
	publisher    = {Springer},
	volume       = {21},
	number       = {6-7},
	pages        = {467--488},
	url          = {https://doi.org/10.1007/BF02650179}
}

@inproceedings{Gily_n_2019,
	title        = {Quantum singular value transformation and beyond: exponential improvements for quantum matrix arithmetics},
	author       = {Gilyén, András and Su, Yuan and Low, Guang Hao and Wiebe, Nathan},
	year         = {2019},
	month        = {June},
	booktitle    = {Proceedings of the 51st Annual ACM SIGACT Symposium on Theory of Computing},
	publisher    = {ACM},
	series       = {STOC ’19},
	pages        = {193–204},
	doi          = {10.1145/3313276.3316366},
	url          = {http://dx.doi.org/10.1145/3313276.3316366},
	collection   = {STOC ’19}
}

@book{goodfellow2016deeplearning,
	title        = {Deep Learning},
	author       = {Ian Goodfellow and Yoshua Bengio and Aaron Courville},
	year         = {2016},
	publisher    = {MIT Press},
	note         = {\url{http://www.deeplearningbook.org}}
}

@article{Gyurik2022towardsquantum,
	title        = {Towards quantum advantage via topological data analysis},
	author       = {Gyurik, Casper and Cade, Chris and Dunjko, Vedran},
	year         = {2022},
	month        = nov,
	journal      = {{Quantum}},
	publisher    = {{Verein zur F{\"{o}}rderung des Open Access Publizierens in den Quantenwissenschaften}},
	volume       = {6},
	pages        = {855},
	doi          = {10.22331/q-2022-11-10-855},
	issn         = {2521-327X},
	url          = {https://doi.org/10.22331/q-2022-11-10-855}
}

@article{gyurik2026provable,
	title        = {Provable Quantum Speedups for Computing Persistence in Topological Data Analysis},
	author       = {Gyurik, Casper and Schmidhuber, Alexander and King, Robbie and Dunjko, Vedran and Hayakawa, Ryu},
	year         = {2026},
	month        = {June},
	journal      = {PRX Quantum},
	publisher    = {American Physical Society (APS)},
	volume       = {7},
	number       = {2},
	doi          = {10.1103/gvys-hl8h},
	issn         = {2691-3399},
	url          = {http://dx.doi.org/10.1103/gvys-hl8h}
}

@article{Havl19,
	title        = {Supervised learning with quantum-enhanced feature spaces},
	author       = {Havl{\'\i}{\v c}ek, Vojt{\v e}ch and C{\'o}rcoles, Antonio D. and Temme, Kristan and Harrow, Aram W. and Kandala, Abhinav and Chow, Jerry M. and Gambetta, Jay M.},
	year         = {2019},
	journal      = {Nature},
	volume       = {567},
	number       = {7747},
	pages        = {209--212},
	doi          = {10.1038/s41586-019-0980-2},
	isbn         = {1476-4687},
	url          = {https://doi.org/10.1038/s41586-019-0980-2},
	date         = {2019/03/01},
	id           = {Havl{\'\i}{\v c}ek2019}
}

@article{hayakawa2022quantumalgorithm,
	title        = {Quantum algorithm for persistent {B}etti numbers and topological data analysis},
	author       = {Hayakawa, Ryu},
	year         = {2022},
	month        = dec,
	journal      = {{Quantum}},
	publisher    = {{Verein zur F{\"{o}}rderung des Open Access Publizierens in den Quantenwissenschaften}},
	volume       = {6},
	pages        = {873},
	doi          = {10.22331/q-2022-12-07-873},
	issn         = {2521-327X},
	url          = {https://doi.org/10.22331/q-2022-12-07-873}
}

@article{Hayakawa2026quantumwalks,
	title        = {Quantum {W}alks on {S}implicial {C}omplexes and {H}armonic {H}omology: {A}pplication to {T}opological {D}ata {A}nalysis with {S}uperpolynomial {S}peedups},
	author       = {Hayakawa, Ryu and Chen, Kuo-Chin and Hsieh, Min-Hsiu},
	year         = {2026},
	month        = jun,
	journal      = {{Quantum}},
	publisher    = {{Verein zur F{\"{o}}rderung des Open Access Publizierens in den Quantenwissenschaften}},
	volume       = {10},
	pages        = {2138},
	doi          = {10.22331/q-2026-06-15-2138},
	issn         = {2521-327X},
	url          = {https://doi.org/10.22331/q-2026-06-15-2138}
}

@article{hur2024neuralquantumembedding,
	title        = {Neural quantum embedding: Pushing the limits of quantum supervised learning},
	author       = {Hur, Tak and Araujo, Israel F. and Park, Daniel K.},
	year         = {2024},
	month        = {Aug},
	journal      = {Phys. Rev. A},
	publisher    = {American Physical Society},
	volume       = {110},
	pages        = {022411},
	doi          = {10.1103/PhysRevA.110.022411},
	url          = {https://link.aps.org/doi/10.1103/PhysRevA.110.022411},
	issue        = {2},
	numpages     = {17}
}

@inproceedings{incudini2023higherorder,
	title        = {Higher-Order Topological Kernels via Quantum Computation},
	author       = {Incudini, Massimiliano and Martini, Francesco and Di Pierro, Alessandra},
	year         = {2023},
	booktitle    = {2023 IEEE International Conference on Quantum Computing and Engineering (QCE)},
	volume       = {01},
	number       = {},
	pages        = {621--629},
	doi          = {10.1109/QCE57702.2023.00076}
}

@article{jager2026quantumfeaturemap,
	title        = {Quantum feature-map learning with reduced resource overhead},
	author       = {J\"ager, Jonas and Els\"asser, Philipp and Torabian, Elham},
	year         = {2026},
	month        = {Jun},
	journal      = {Phys. Rev. Res.},
	publisher    = {American Physical Society},
	volume       = {8},
	pages        = {023247},
	doi          = {10.1103/v29j-rh32},
	url          = {https://link.aps.org/doi/10.1103/v29j-rh32},
	issue        = {2},
	numpages     = {23}
}

@misc{kerenidis2022quantummachinelearningsubspace,
	title        = {Quantum machine learning with subspace states},
	author       = {Iordanis Kerenidis and Anupam Prakash},
	year         = {2022},
	url          = {https://arxiv.org/abs/2202.00054},
	eprint       = {2202.00054},
	archiveprefix = {arXiv},
	primaryclass = {quant-ph}
}

@article{LaRose_2020,
	title        = {Robust data encodings for quantum classifiers},
	author       = {LaRose, Ryan and Coyle, Brian},
	year         = {2020},
	month        = {Sept},
	journal      = {Physical Review A},
	publisher    = {American Physical Society (APS)},
	volume       = {102},
	number       = {3},
	issn         = {2469-9934},
	url          = {http://dx.doi.org/10.1103/PhysRevA.102.032420}
}

@article{LCU,
	title        = {Hamiltonian simulation using linear combinations of unitary operations},
	author       = {Childs, Andrew M. and Wiebe, Nathan},
	year         = {2012},
	month        = nov,
	journal      = {Quantum Info. Comput.},
	publisher    = {Rinton Press, Incorporated},
	address      = {Paramus, NJ},
	volume       = {12},
	number       = {11–12},
	pages        = {901–924},
	doi          = {10.26421/QIC12.11-12-1},
	issn         = {1533-7146},
	url          = {https://doi.org/10.26421/QIC12.11-12-1},
	issue_date   = {November 2012},
	numpages     = {24}
}

@inproceedings{LCUberry2015hamiltonian,
	title        = {Hamiltonian simulation with nearly optimal dependence on all parameters},
	author       = {Berry, Dominic W and Childs, Andrew M and Kothari, Robin},
	year         = {2015},
	booktitle    = {2015 IEEE 56th annual symposium on foundations of computer science},
	pages        = {792--809},
	doi          = {10.1109/FOCS.2015.54},
	url          = {https://doi.org/10.1109/FOCS.2015.54},
	organization = {IEEE}
}

@article{le2025optimizing,
	title        = {Optimizing quantum convolutional neural network architectures for arbitrary data dimension},
	author       = {Lee, Changwon  and Araujo, Israel F.  and Kim, Dongha  and Lee, Junghan  and Park, Siheon  and Ryu, Ju-Young  and Park, Daniel K.},
	year         = {2025},
	journal      = {Frontiers in Physics},
	volume       = {13},
	doi          = {10.3389/fphy.2025.1529188},
	issn         = {2296-424X},
	url          = {https://www.frontiersin.org/journals/physics/articles/10.3389/fphy.2025.1529188}
}

@article{leykam2023topological,
	title        = {Topological data analysis and machine learning},
	author       = {Leykam, Daniel and Angelakis, Dimitris G},
	year         = {2023},
	journal      = {Advances in Physics: X},
	publisher    = {Taylor \& Francis},
	volume       = {8},
	number       = {1},
	pages        = {2202331},
	url          = {https://doi.org/10.1080/23746149.2023.2202331}
}

@article{lloyd1996universal,
	title        = {Universal quantum simulators},
	author       = {Lloyd, Seth},
	year         = {1996},
	journal      = {Science},
	publisher    = {American Association for the Advancement of Science},
	volume       = {273},
	number       = {5278},
	pages        = {1073--1078},
	doi          = {10.1126/science.273.5278.1073},
	url          = {https://doi.org}
}

@article{lloyd2016quantum,
	title        = {Quantum algorithms for topological and geometric analysis of data},
	author       = {Lloyd, Seth and Garnerone, Silvano and Zanardi, Paolo},
	year         = {2016},
	journal      = {Nature Communications},
	publisher    = {Nature Publishing Group UK London},
	volume       = {7},
	number       = {1},
	pages        = {10138}
}

@misc{lloyd2020quantumembeddingsmachinelearning,
	title        = {Quantum embeddings for machine learning},
	author       = {Seth Lloyd and Maria Schuld and Aroosa Ijaz and Josh Izaac and Nathan Killoran},
	year         = {2020},
	url          = {https://arxiv.org/abs/2001.03622},
	eprint       = {2001.03622},
	archiveprefix = {arXiv},
	primaryclass = {quant-ph}
}

@article{Low_2019,
	title        = {Hamiltonian Simulation by Qubitization},
	author       = {Low, Guang Hao and Chuang, Isaac L.},
	year         = {2019},
	month        = {07},
	journal      = {Quantum},
	publisher    = {Verein zur Forderung des Open Access Publizierens in den Quantenwissenschaften},
	volume       = {3},
	pages        = {163},
	doi          = {10.22331/q-2019-07-12-163},
	issn         = {2521-327X},
	url          = {http://dx.doi.org/10.22331/q-2019-07-12-163}
}

@article{lum2013extractingwithtda,
	title        = {Extracting insights from the shape of complex data using topology},
	author       = {Lum, Pek Y and Singh, Gurjeet and Lehman, Alan and Ishkanov, Tigran and Vejdemo-Johansson, Mikael and Alagappan, Muthu and Carlsson, John and Carlsson, Gunnar},
	year         = {2013},
	journal      = {Scientific Reports},
	publisher    = {Nature Publishing Group UK London},
	volume       = {3},
	number       = {1},
	pages        = {1236},
	doi          = {10.1038/srep01236},
	url          = {https://doi.org/10.1038/srep01236}
}

@book{manin2007mathematics,
	title        = {Mathematics as Metaphor: Selected Essays of Yuri I. Manin},
	author       = {Manin, I.U.I.},
	year         = {2007},
	publisher    = {American Mathematical Society},
	series       = {[Collected works (American Mathematical Society)},
	isbn         = {9780821843314}
}

@article{marshall2023highdimensional,
	title        = {High {D}imensional {Q}uantum {M}achine {L}earning {W}ith {S}mall {Q}uantum {C}omputers},
	author       = {Marshall, Simon C. and Gyurik, Casper and Dunjko, Vedran},
	year         = {2023},
	month        = aug,
	journal      = {{Quantum}},
	publisher    = {{Verein zur F{\"{o}}rderung des Open Access Publizierens in den Quantenwissenschaften}},
	volume       = {7},
	pages        = {1078},
	doi          = {10.22331/q-2023-08-09-1078},
	issn         = {2521-327X},
	url          = {https://doi.org/10.22331/q-2023-08-09-1078}
}

@article{mcardle2026streamlinedquantum,
	title        = {A streamlined quantum algorithm for topological data analysis with exponentially fewer qubits},
	author       = {McArdle, Sam and Gily{\'{e}}n,, Andr{\'{a}}s and Berta, Mario},
	year         = {2026},
	month        = apr,
	journal      = {{Quantum}},
	publisher    = {{Verein zur F{\"{o}}rderung des Open Access Publizierens in den Quantenwissenschaften}},
	volume       = {10},
	pages        = {2058},
	doi          = {10.22331/q-2026-04-10-2058},
	issn         = {2521-327X},
	url          = {https://doi.org/10.22331/q-2026-04-10-2058}
}

@article{mengoni2019persistent,
	title        = {Persistent homology analysis of multiqubit entanglement},
	author       = {Riccardo Mengoni and Alessandra Di Pierro and Laleh Memarzadeh and Stefano Mancini},
	year         = {2019},
	journal      = {Quantum Inf. Comput.},
	volume       = {20},
	pages        = {375--399},
	url          = {https://api.semanticscholar.org/CorpusID:196831331}
}

@book{murphy2012machine,
	title        = {Machine learning: a probabilistic perspective},
	author       = {Murphy, Kevin P},
	year         = {2012},
	publisher    = {MIT press},
	isbn         = {978-0-262-01802-9},
	url          = {https://mitpress.mit.edu/9780262018029/machine-learning/}
}

@misc{nghiem2025quantumtopologicaldataanalysis,
	title        = {Towards quantum topological data analysis: torsion detection},
	author       = {Nhat A. Nghiem},
	year         = {2025},
	url          = {https://arxiv.org/abs/2508.19943},
	eprint       = {2508.19943},
	archiveprefix = {arXiv},
	primaryclass = {quant-ph}
}

@article{nicolau2011topology,
	title        = {Topology based data analysis identifies a subgroup of breast cancers with a unique mutational profile and excellent survival},
	author       = {Nicolau, Monica and Levine, Arnold J and Carlsson, Gunnar},
	year         = {2011},
	journal      = {Proceedings of the National Academy of Sciences},
	publisher    = {National Academy of Sciences},
	volume       = {108},
	number       = {17},
	pages        = {7265--7270},
	doi          = {10.1073/pnas.1102826108},
	url          = {https://doi.org/10.1073/pnas.1102826108}
}

@article{Park_Light_1986Krylov,
	title        = {{Unitary quantum time evolution by iterative Lanczos reduction}},
	author       = {Park, Tae Jun and Light, John C.},
	year         = {1986},
	journal      = {The Journal of Chemical Physics},
	volume       = {85},
	number       = {10},
	pages        = {5870--5876},
	doi          = {10.1063/1.451548},
	url          = {https://pubs.aip.org/aip/jcp/article/85/10/5870/94677/Unitary-quantum-time-evolution-by-iterative}
}

@article{perezsalinas2020datareuploading,
	title        = {Data re-uploading for a universal quantum classifier},
	author       = {P{\'{e}}rez-Salinas, Adri{\'{a}}n and Cervera-Lierta, Alba and Gil-Fuster, Elies and Latorre, Jos{\'{e}} I.},
	year         = {2020},
	month        = feb,
	journal      = {{Quantum}},
	publisher    = {{Verein zur F{\"{o}}rderung des Open Access Publizierens in den Quantenwissenschaften}},
	volume       = {4},
	pages        = {226},
	doi          = {10.22331/q-2020-02-06-226},
	issn         = {2521-327X},
	url          = {https://doi.org/10.22331/q-2020-02-06-226}
}

@article{PhysRevA.104.032416,
	title        = {Quantum evolution kernel: Machine learning on graphs with programmable arrays of qubits},
	author       = {Henry, Louis-Paul and Thabet, Slimane and Dalyac, Constantin and Henriet, Lo\"{\i}c},
	year         = {2021},
	month        = {Sep},
	journal      = {Phys. Rev. A},
	publisher    = {American Physical Society},
	volume       = {104},
	pages        = {032416},
	doi          = {10.1103/PhysRevA.104.032416},
	url          = {https://link.aps.org/doi/10.1103/PhysRevA.104.032416},
	issue        = {3},
	numpages     = {16}
}

@article{pun2022persistent,
	title        = {Persistent-homology-based machine learning: a survey and a comparative study},
	author       = {Pun, Chi Seng and Lee, Si Xian and Xia, Kelin},
	year         = {2022},
	journal      = {Artificial Intelligence Review},
	publisher    = {Springer},
	volume       = {55},
	number       = {7},
	pages        = {5169--5213},
	url          = {https://doi.org/10.1007/s10462-022-10146-z}
}

@article{rabadan2020identification,
	title        = {Identification of relevant genetic alterations in cancer using topological data analysis},
	author       = {Rabad{\'a}n, Ra{\'u}l and Mohamedi, Yamina and Rubin, Udi and Chu, Tim and Alghalith, Adam N and Elliott, Oliver and Arn{\'e}s, Luis and Cal, Santiago and Obaya, {\'A}lvaro J and Levine, Arnold J and others},
	year         = {2020},
	journal      = {Nature Communications},
	publisher    = {Nature Publishing Group UK London},
	volume       = {11},
	number       = {1},
	pages        = {3808},
	doi          = {10.1038/s41467-020-17659-7},
	url          = {https://doi.org}
}

@article{Scali_2024,
	title        = {Quantum topological data analysis via the estimation of the density of states},
	author       = {Scali, Stefano and Umeano, Chukwudubem and Kyriienko, Oleksandr},
	year         = {2024},
	month        = {Oct},
	journal      = {Physical Review A},
	publisher    = {American Physical Society (APS)},
	volume       = {110},
	number       = {4},
	issn         = {2469-9934},
	url          = {http://dx.doi.org/10.1103/PhysRevA.110.042616}
}

@article{Schuld_2019,
	title        = {Quantum Machine Learning in Feature {H}ilbert Spaces},
	author       = {Schuld, Maria and Killoran, Nathan},
	year         = {2019},
	month        = {Feb},
	journal      = {Physical Review Letters},
	publisher    = {American Physical Society (APS)},
	volume       = {122},
	number       = {4},
	issn         = {1079-7114},
	url          = {http://dx.doi.org/10.1103/PhysRevLett.122.040504}
}

@article{schuld2021effect,
	title        = {The effect of data encoding on the expressive power of variational quantum-machine-learning models},
	author       = {Schuld, Maria and Sweke, Ryan and Meyer, Johannes Jakob},
	year         = {2021},
	journal      = {Physical Review A},
	volume       = {103},
	number       = {3},
	pages        = {032430},
	doi          = {10.1103/PhysRevA.103.032430},
	eprint       = {2008.08605},
	archiveprefix = {arXiv},
	primaryclass = {quant-ph}
}

@book{schuld2021machine,
	title        = {Machine Learning with Quantum Computers},
	author       = {Schuld, M. and Petruccione, F.},
	year         = {2021},
	publisher    = {Springer International Publishing},
	series       = {Quantum Science and Technology},
	isbn         = {9783030830984},
	url          = {https://books.google.com.sg/books?id=-N5IEAAAQBAJ}
}

@misc{schuld2021supervisedquantummachinelearning,
	title        = {Supervised quantum machine learning models are kernel methods},
	author       = {Maria Schuld},
	year         = {2021},
	url          = {https://arxiv.org/abs/2101.11020},
	eprint       = {2101.11020},
	archiveprefix = {arXiv},
	primaryclass = {quant-ph}
}

@article{shin2023exponential,
	title        = {Exponential data encoding for quantum supervised learning},
	author       = {Shin, S. and Teo, Y. S. and Jeong, H.},
	year         = {2023},
	month        = {Jan},
	journal      = {Phys. Rev. A},
	publisher    = {American Physical Society},
	volume       = {107},
	pages        = {012422},
	doi          = {10.1103/PhysRevA.107.012422},
	url          = {https://link.aps.org/doi/10.1103/PhysRevA.107.012422},
	issue        = {1},
	numpages     = {20}
}

@article{singh2023topological,
	title        = {Topological data analysis in medical imaging: current state of the art},
	author       = {Singh, Yashbir and Farrelly, Colleen M and Hathaway, Quincy A and Leiner, Tim and Jagtap, Jaidip and Carlsson, Gunnar E and Erickson, Bradley J},
	year         = {2023},
	journal      = {Insights into Imaging},
	publisher    = {Springer},
	volume       = {14},
	number       = {1},
	pages        = {58},
	doi          = {10.1186/s13244-023-01413-w},
	url          = {https://doi.org/10.1186/s13244-023-01413-w}
}

@misc{tanner2026nonvariationalsupervisedquantumkernel,
	title        = {Non-variational supervised quantum kernel methods: a review},
	author       = {John Tanner and Chon-Fai Kam and Jingbo Wang},
	year         = {2026},
	url          = {https://arxiv.org/abs/2604.07896},
	eprint       = {2604.07896},
	archiveprefix = {arXiv},
	primaryclass = {quant-ph}
}

@article{topaz2015topological,
	title        = {Topological data analysis of biological aggregation models},
	author       = {Topaz, Chad M and Ziegelmeier, Lori and Halverson, Tom},
	year         = {2015},
	journal      = {PloS one},
	publisher    = {Public Library of Science San Francisco, CA USA},
	volume       = {10},
	number       = {5},
	pages        = {e0126383},
	doi          = {10.1371/journal.pone.0126383},
	url          = {https://doi.org/10.1371/journal.pone.0126383}
}

@misc{ubaru2021quantumtopologicaldataanalysis,
	title        = {Quantum Topological Data Analysis with Linear Depth and Exponential Speedup},
	author       = {Shashanka Ubaru and Ismail Yunus Akhalwaya and Mark S. Squillante and Kenneth L. Clarkson and Lior Horesh},
	year         = {2021},
	url          = {https://arxiv.org/abs/2108.02811},
	eprint       = {2108.02811},
	archiveprefix = {arXiv},
	primaryclass = {quant-ph}
}

@article{wee2023persistent,
	title        = {Persistent {D}irac for molecular representation},
	author       = {Wee, JunJie and Bianconi, Ginestra and Xia, Kelin},
	year         = {2023},
	journal      = {Scientific Reports},
	publisher    = {Nature Publishing Group UK London},
	volume       = {13},
	number       = {1},
	pages        = {11183},
	url          = {https://doi.org/10.1038/s41598-023-37853-z}
}

@article{wu2023novel,
	title        = {A novel approach to topological network analysis for the identification of metrics and signatures in non-small cell lung cancer},
	author       = {Wu, Isabella and Wang, Xin},
	year         = {2023},
	journal      = {Scientific Reports},
	publisher    = {Nature Publishing Group UK London},
	volume       = {13},
	number       = {1},
	pages        = {8223},
	doi          = {10.1038/s41598-023-35165-w},
	url          = {https://doi.org/10.1038/s41598-023-35165-w}
}

@misc{zang2025benchmarkingdataencodingmethods,
	title        = {Benchmarking data encoding methods in Quantum Machine Learning},
	author       = {Orlane Zang and Grégoire Barrué and Tony Quertier},
	year         = {2025},
	url          = {https://arxiv.org/abs/2505.14295},
	eprint       = {2505.14295},
	archiveprefix = {arXiv},
	primaryclass = {quant-ph}
}
\clearpage

\onecolumngrid
\appendix
\section*{Appendices}

\section{Additional similarity matrices}

\begin{figure}[htbp]
  \centering
  \begin{subfigure}[t]{0.32\textwidth}
    \includegraphics[width=\linewidth]{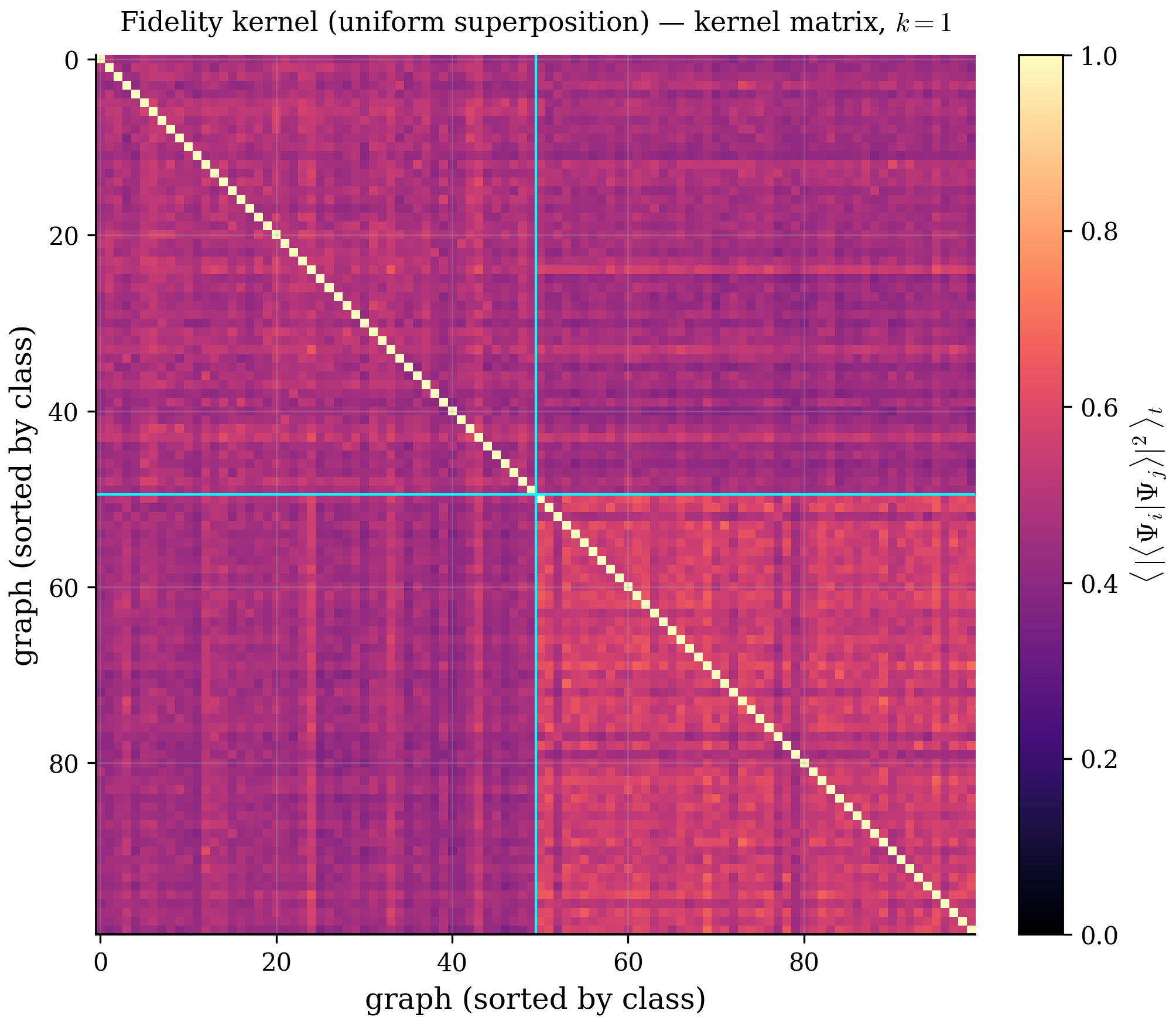}
    \caption{$k=1$}\label{fig:fid-k1}
  \end{subfigure}\hfill
  \begin{subfigure}[t]{0.32\textwidth}
    \includegraphics[width=\linewidth]{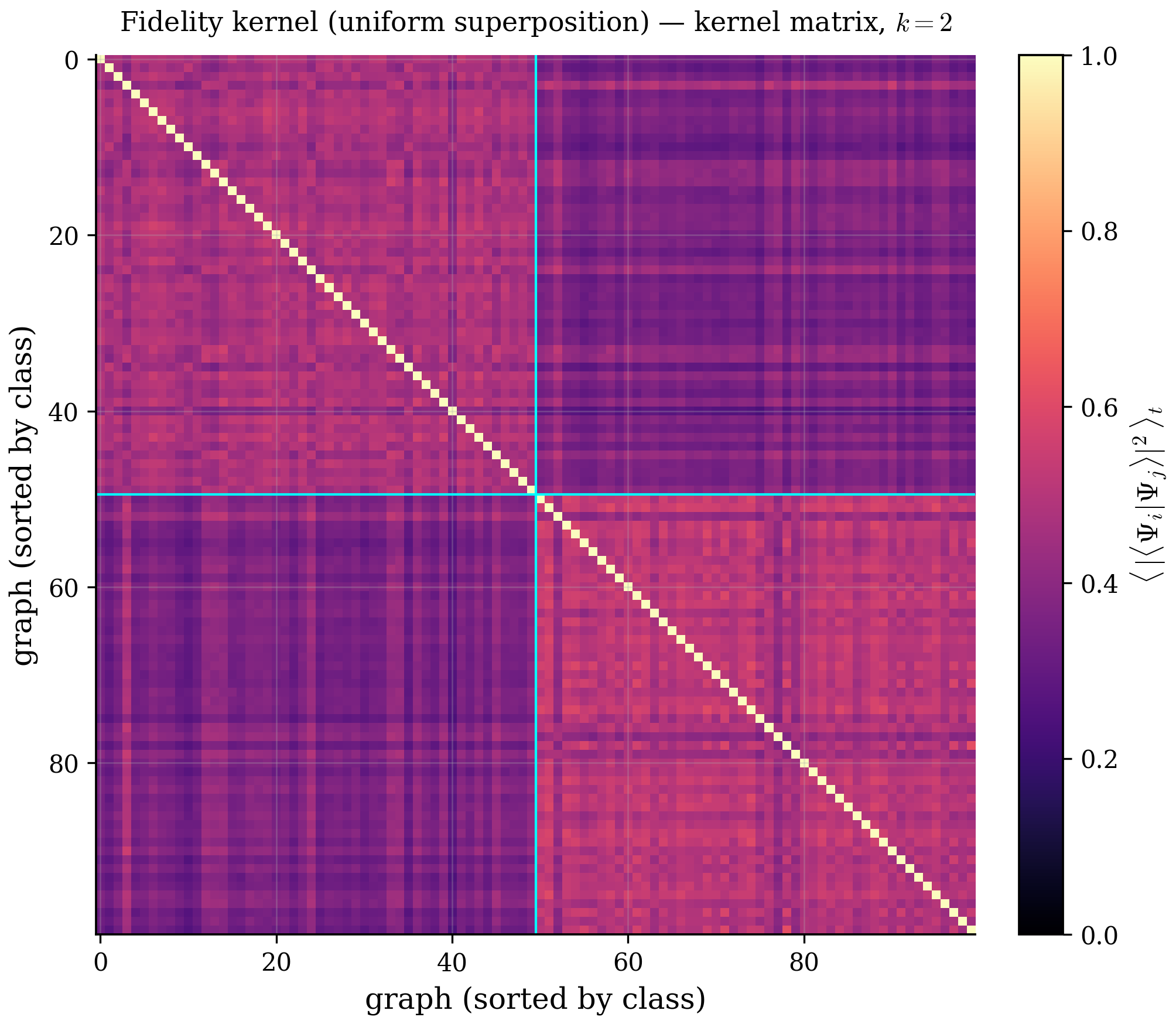}
    \caption{$k=2$}\label{fig:fid-k2}
  \end{subfigure}\hfill
  \begin{subfigure}[t]{0.32\textwidth}
    \includegraphics[width=\linewidth]{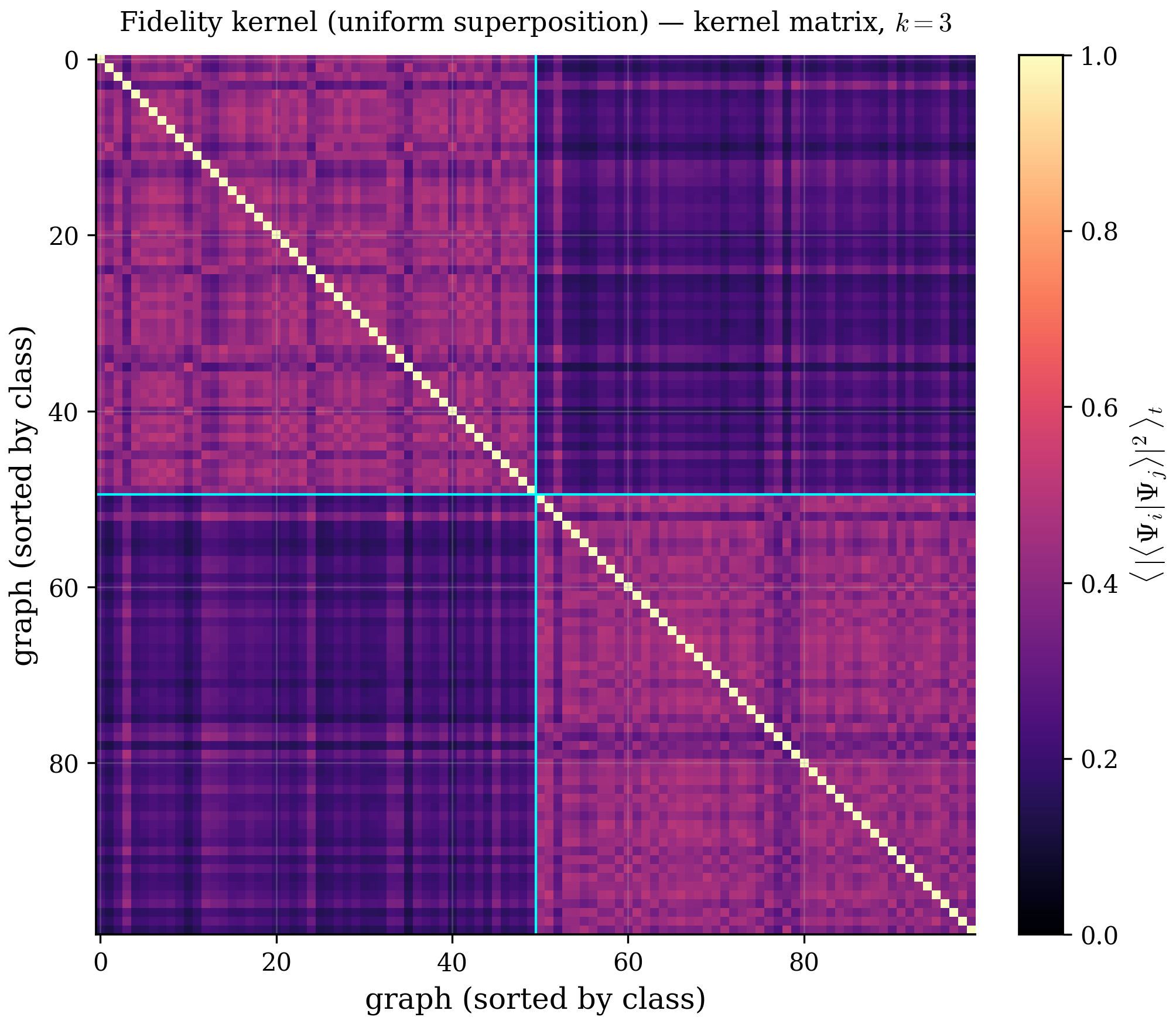}
    \caption{$k=3$}\label{fig:fid-k3}
  \end{subfigure}
 
  \medskip
  \begin{subfigure}[t]{0.32\textwidth}
    \includegraphics[width=\linewidth]{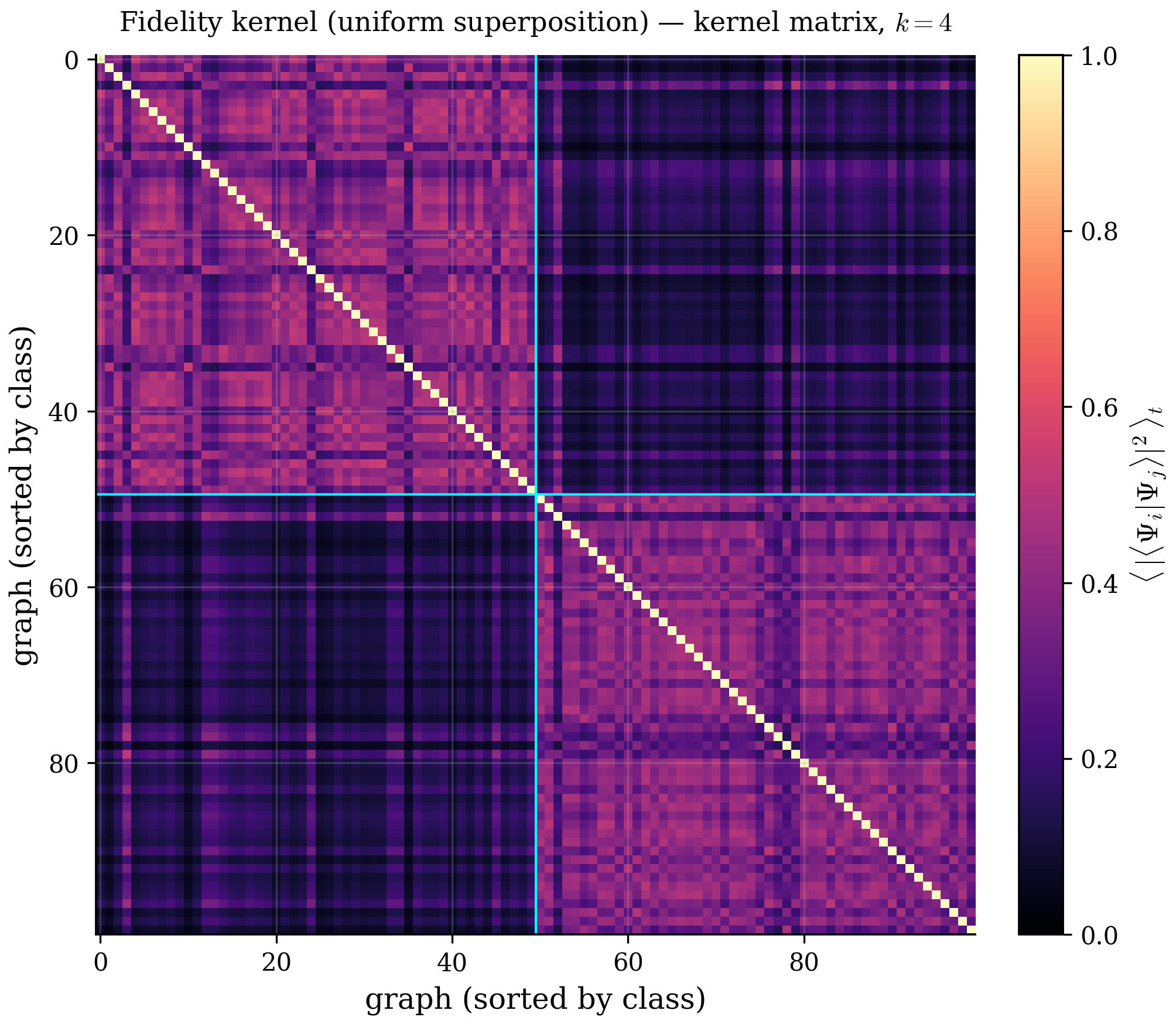}
    \caption{$k=4$}\label{fig:fid-k4}
  \end{subfigure}\hfill
  \begin{subfigure}[t]{0.32\textwidth}
    \includegraphics[width=\linewidth]{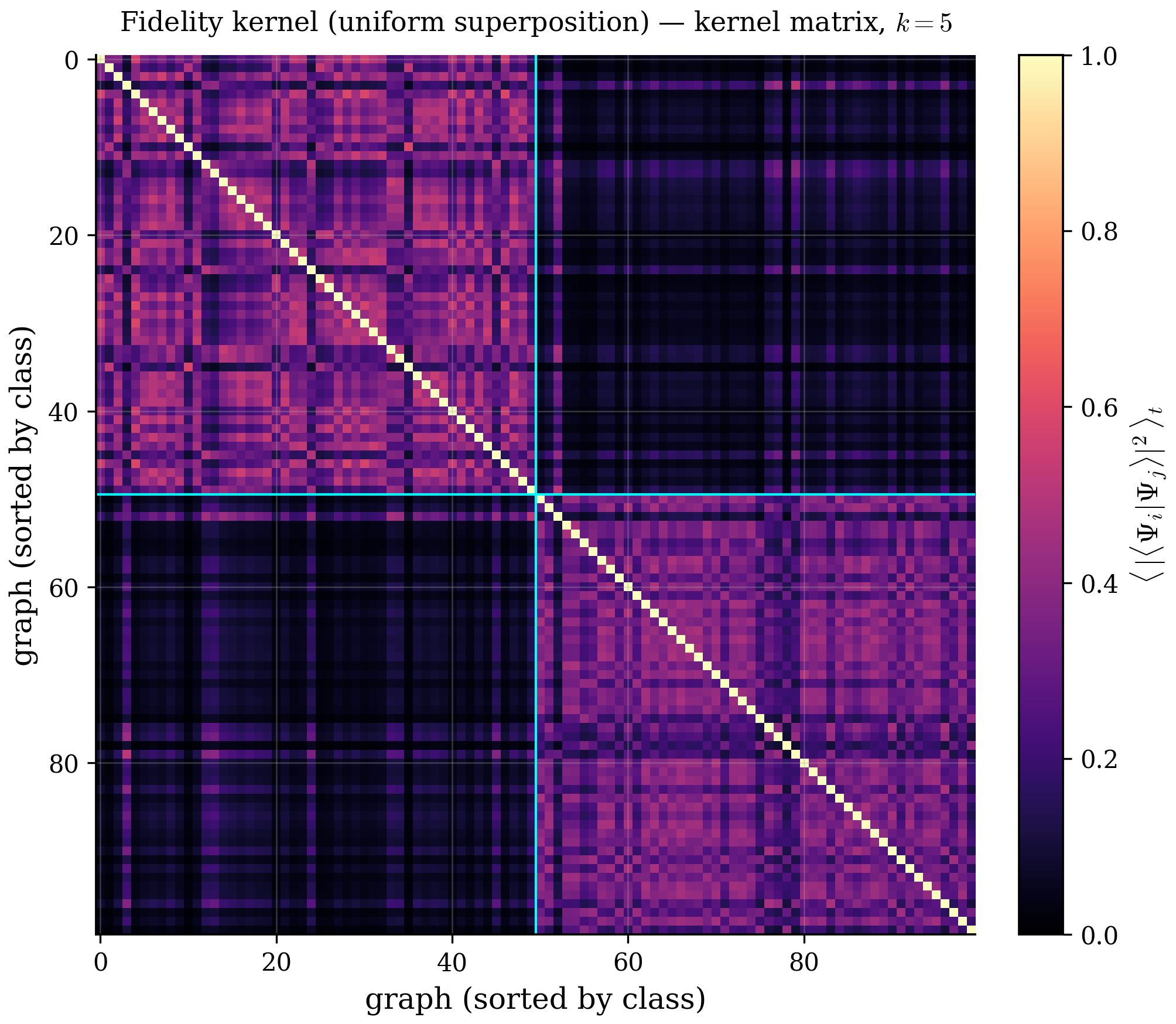}
    \caption{$k=5$}\label{fig:fid-k5}
  \end{subfigure}\hfill
  \begin{subfigure}[t]{0.32\textwidth}
    \includegraphics[width=\linewidth]{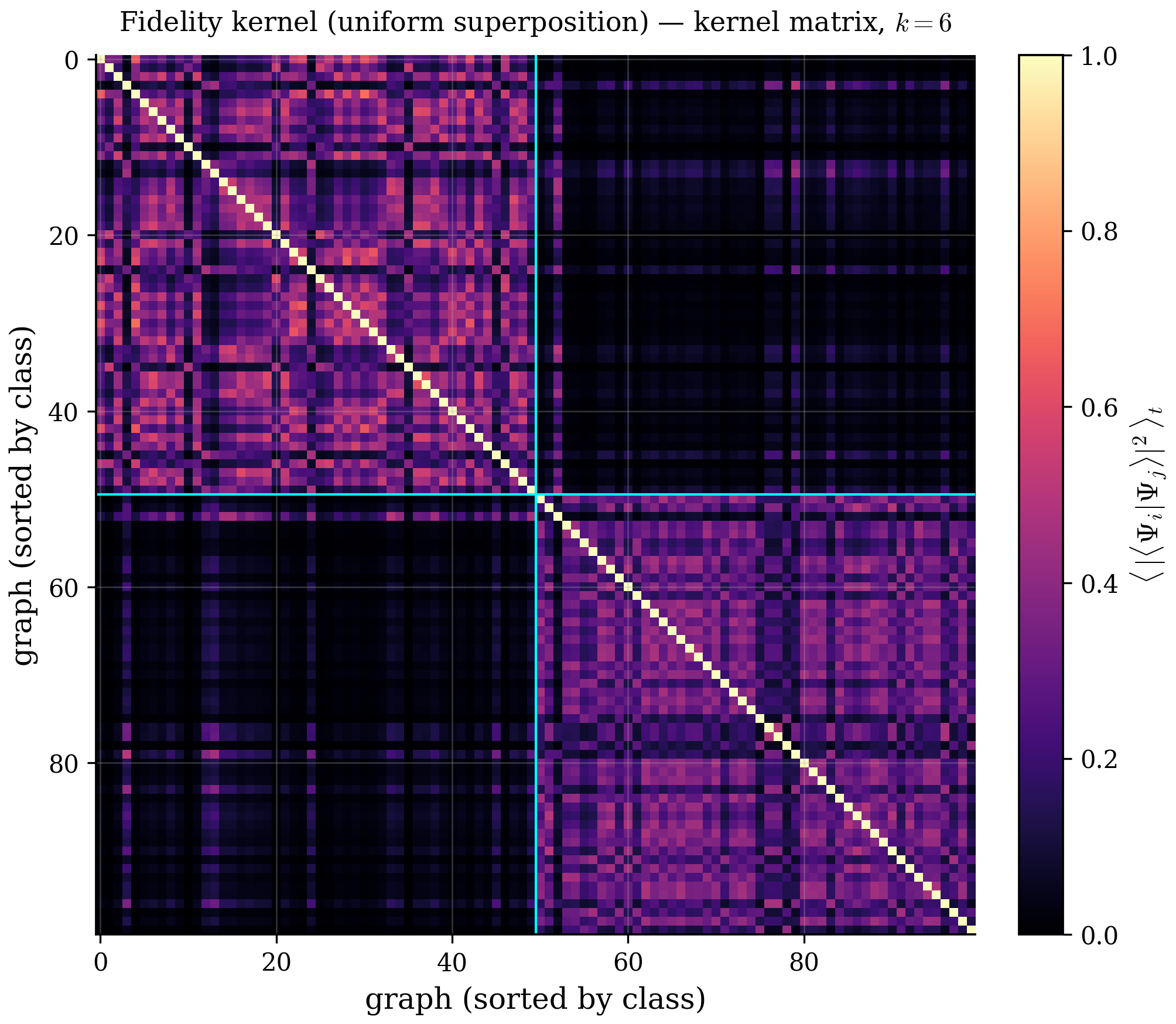}
    \caption{$k=6$}\label{fig:fid-k6}
  \end{subfigure}
 
  \medskip
  \begin{subfigure}[t]{0.32\textwidth}
    \includegraphics[width=\linewidth]{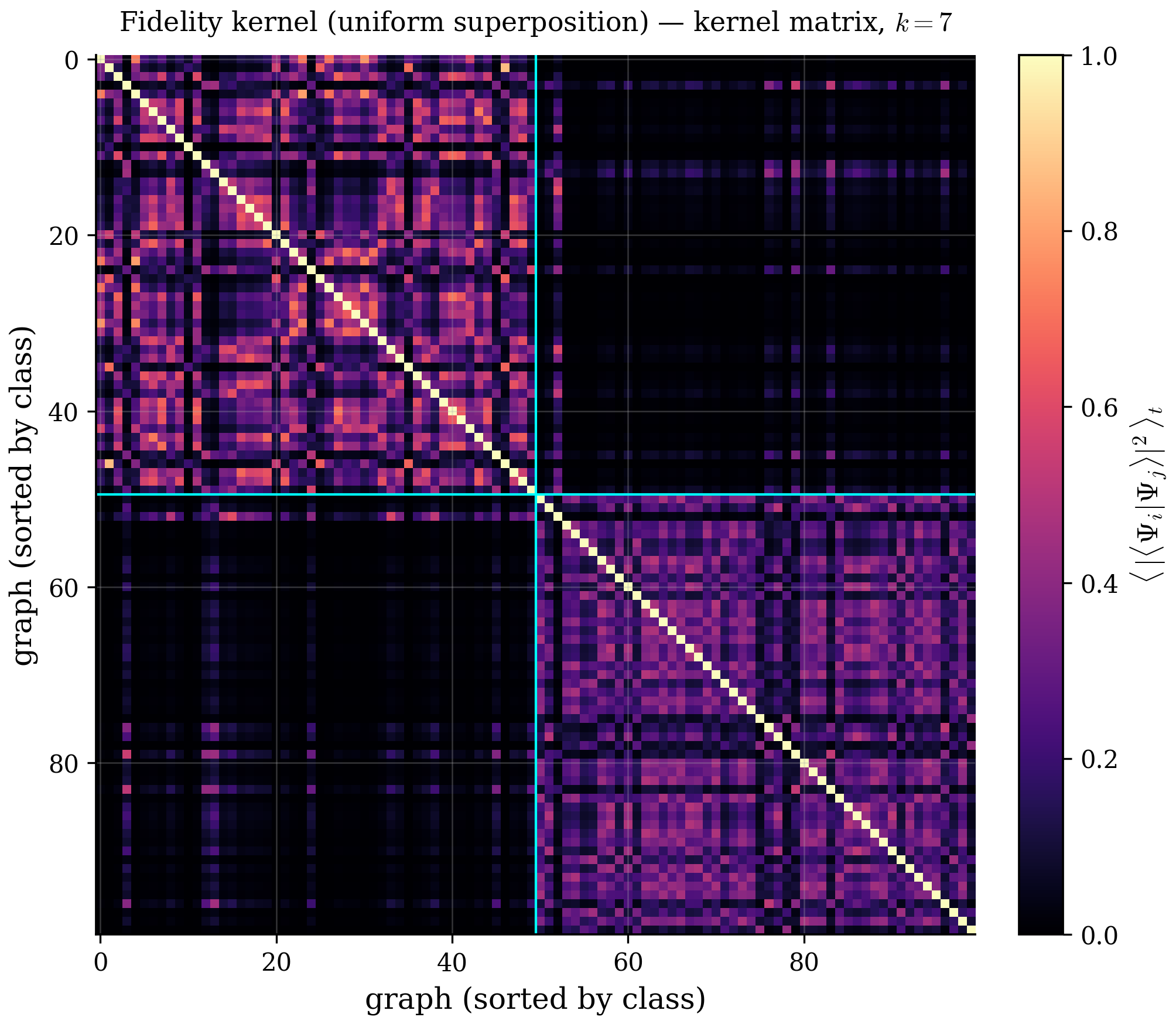}
    \caption{$k=7$}\label{fig:fid-k7}
  \end{subfigure}\hfill
  \begin{subfigure}[t]{0.32\textwidth}
    \includegraphics[width=\linewidth]{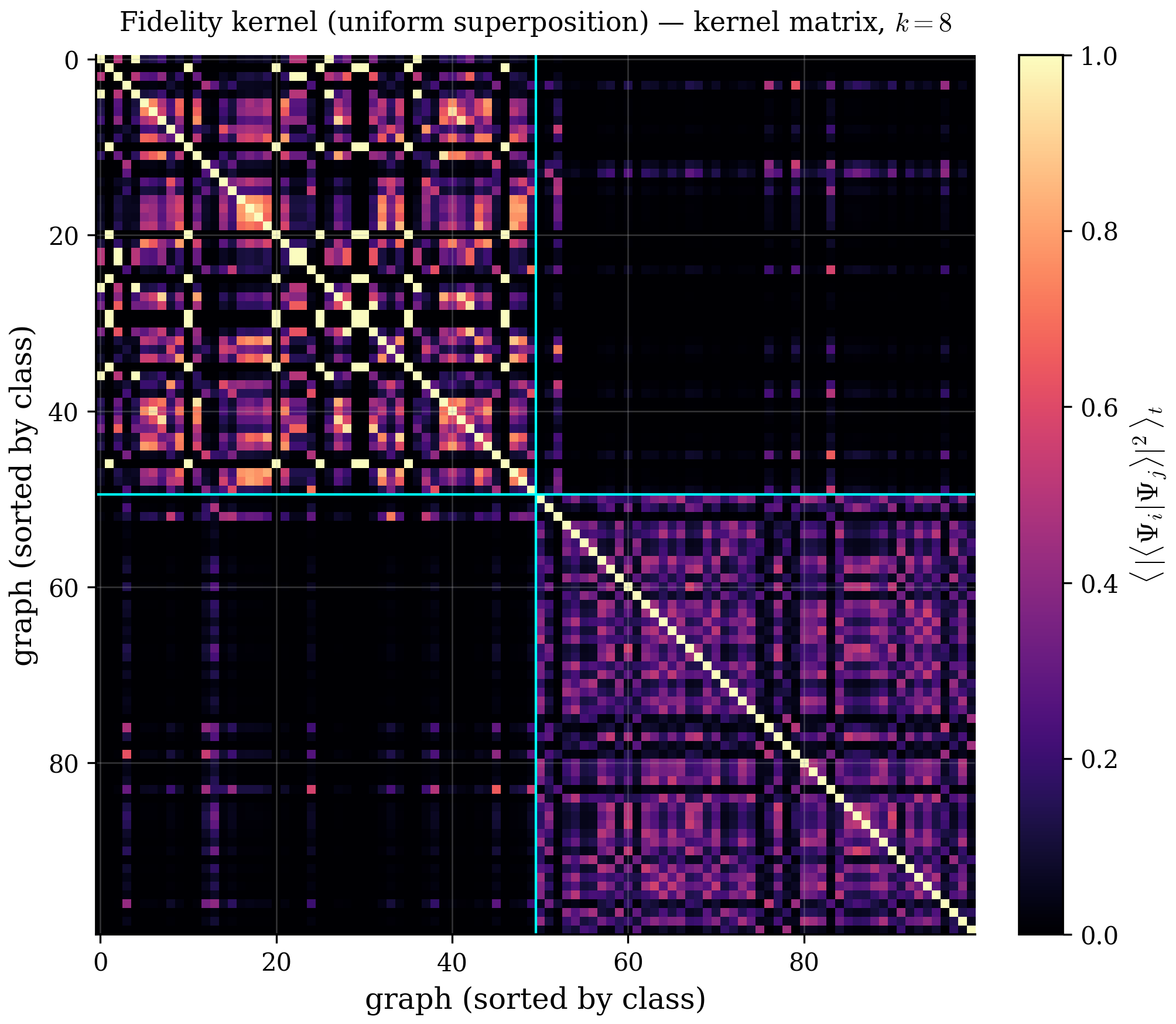}
    \caption{$k=8$}\label{fig:fid-k8}
  \end{subfigure}\hfill
  \begin{subfigure}[t]{0.32\textwidth}
    \includegraphics[width=\linewidth]{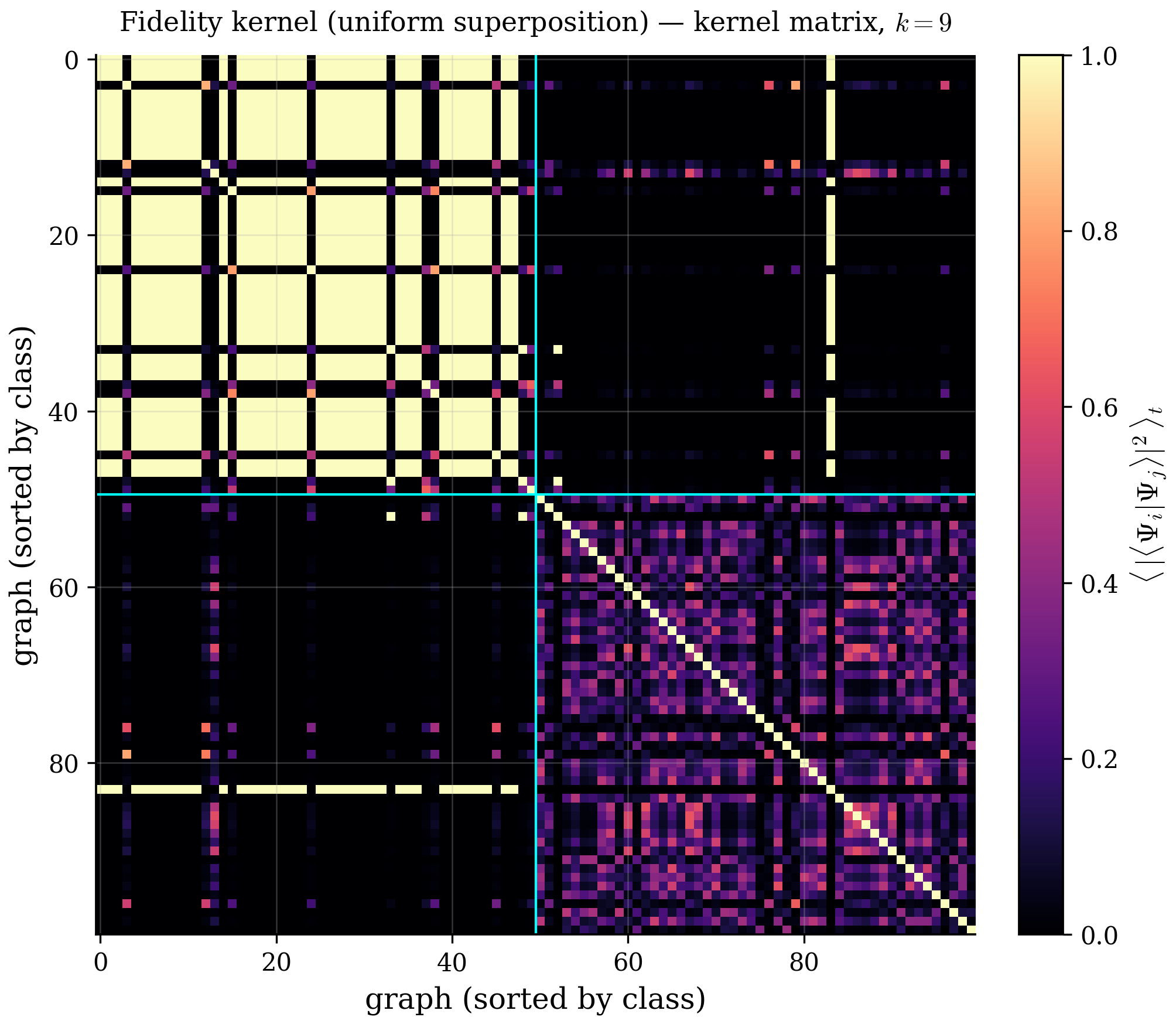}
    \caption{$k=9$}\label{fig:fid-k9}
  \end{subfigure}
 
  \caption{\justifying Fidelity kernel matrices $K_{ij}=\langle|\langle\Psi_i|\Psi_j\rangle|^2\rangle_t$
           for Laplacian dimensions $k=1,\dots,9$, evaluated on $G(50, p)$ with $25$ graphs per class, for class $1$: $p_0=0.5$ and class $2$: $p_1=0.52$.. Rows and columns are sorted by class; the cyan cross marks
           the class boundary.}
  \label{fig:fidelity-grid}
\end{figure}

\begin{figure}[htbp]
  \centering
  \begin{subfigure}[t]{0.32\textwidth}
    \includegraphics[width=\linewidth]{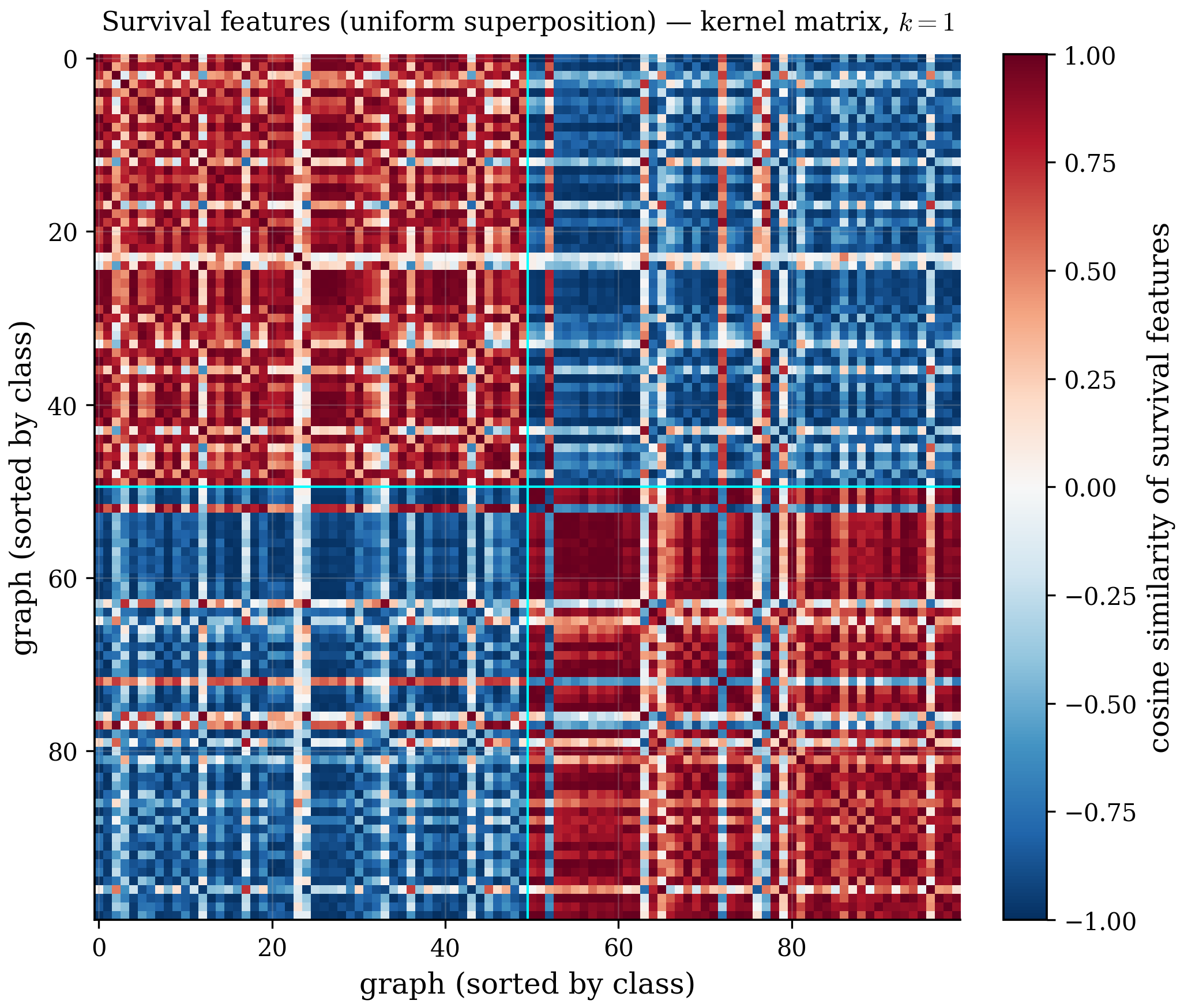}
    \caption{$k=1$}\label{fig:surv-k1}
  \end{subfigure}\hfill
  \begin{subfigure}[t]{0.32\textwidth}
    \includegraphics[width=\linewidth]{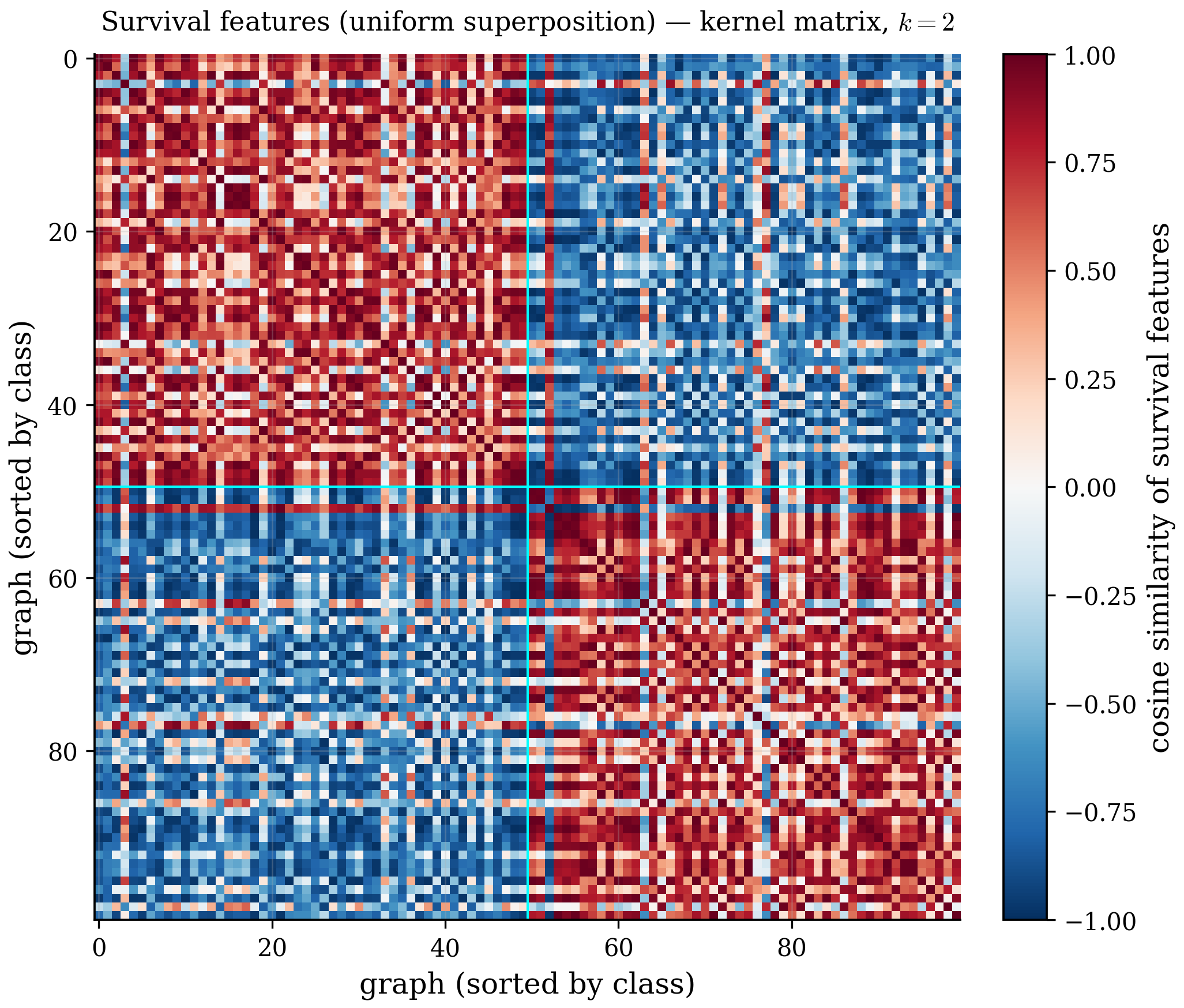}
    \caption{$k=2$}\label{fig:surv-k2}
  \end{subfigure}\hfill
  \begin{subfigure}[t]{0.32\textwidth}
    \includegraphics[width=\linewidth]{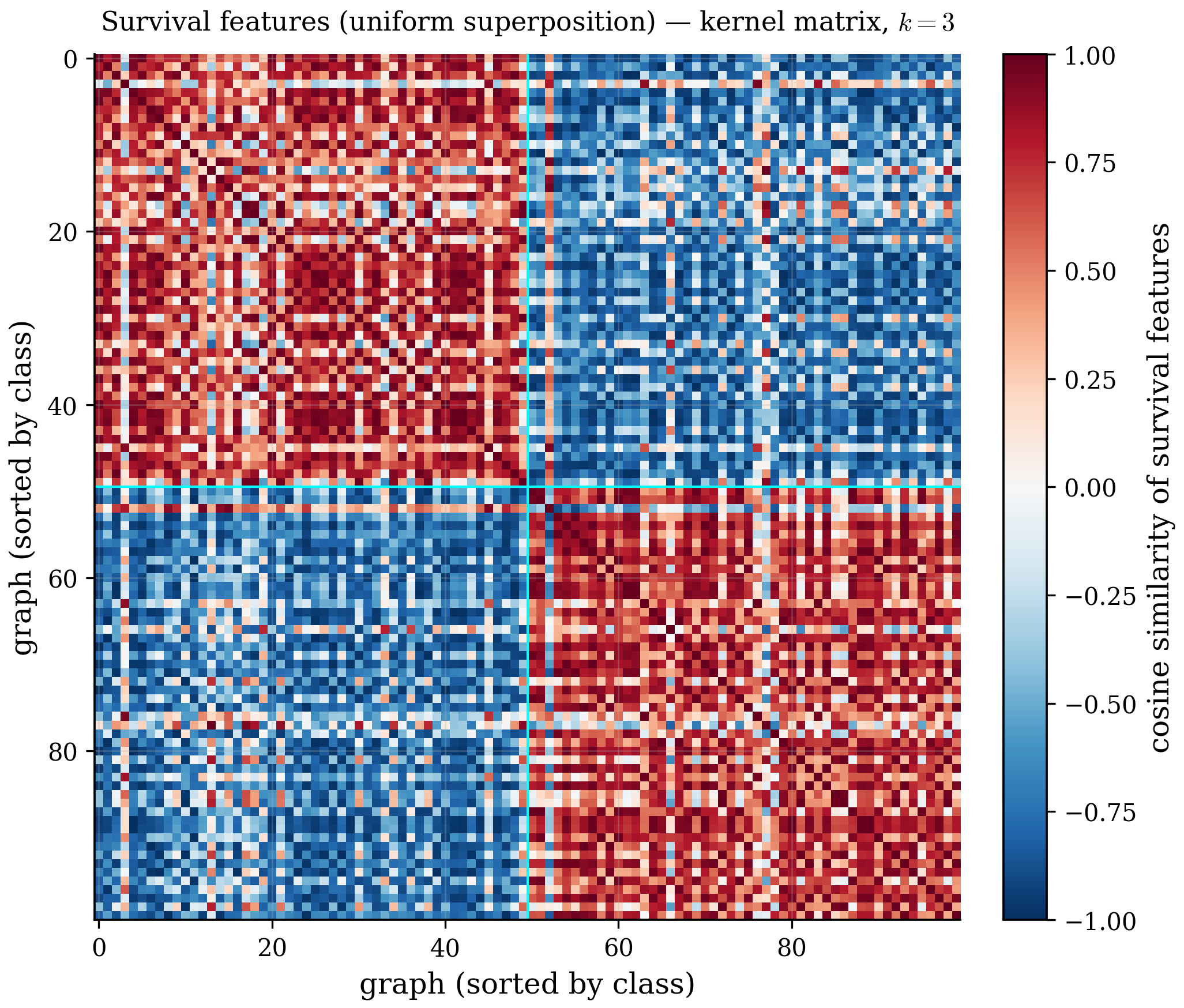}
    \caption{$k=3$}\label{fig:surv-k3}
  \end{subfigure}
 
  \medskip
  \begin{subfigure}[t]{0.32\textwidth}
    \includegraphics[width=\linewidth]{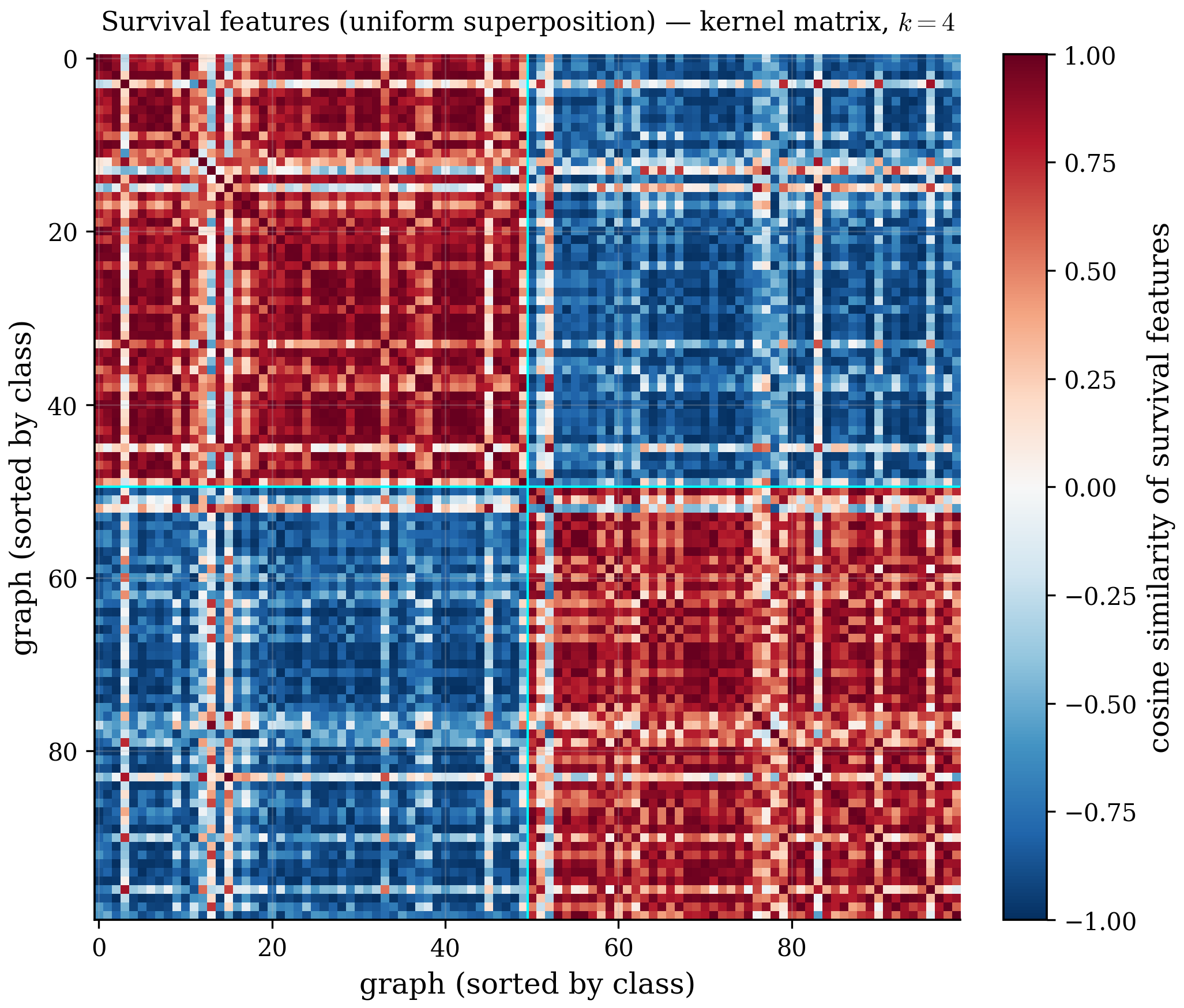}
    \caption{$k=4$}\label{fig:surv-k4}
  \end{subfigure}\hfill
  \begin{subfigure}[t]{0.32\textwidth}
    \includegraphics[width=\linewidth]{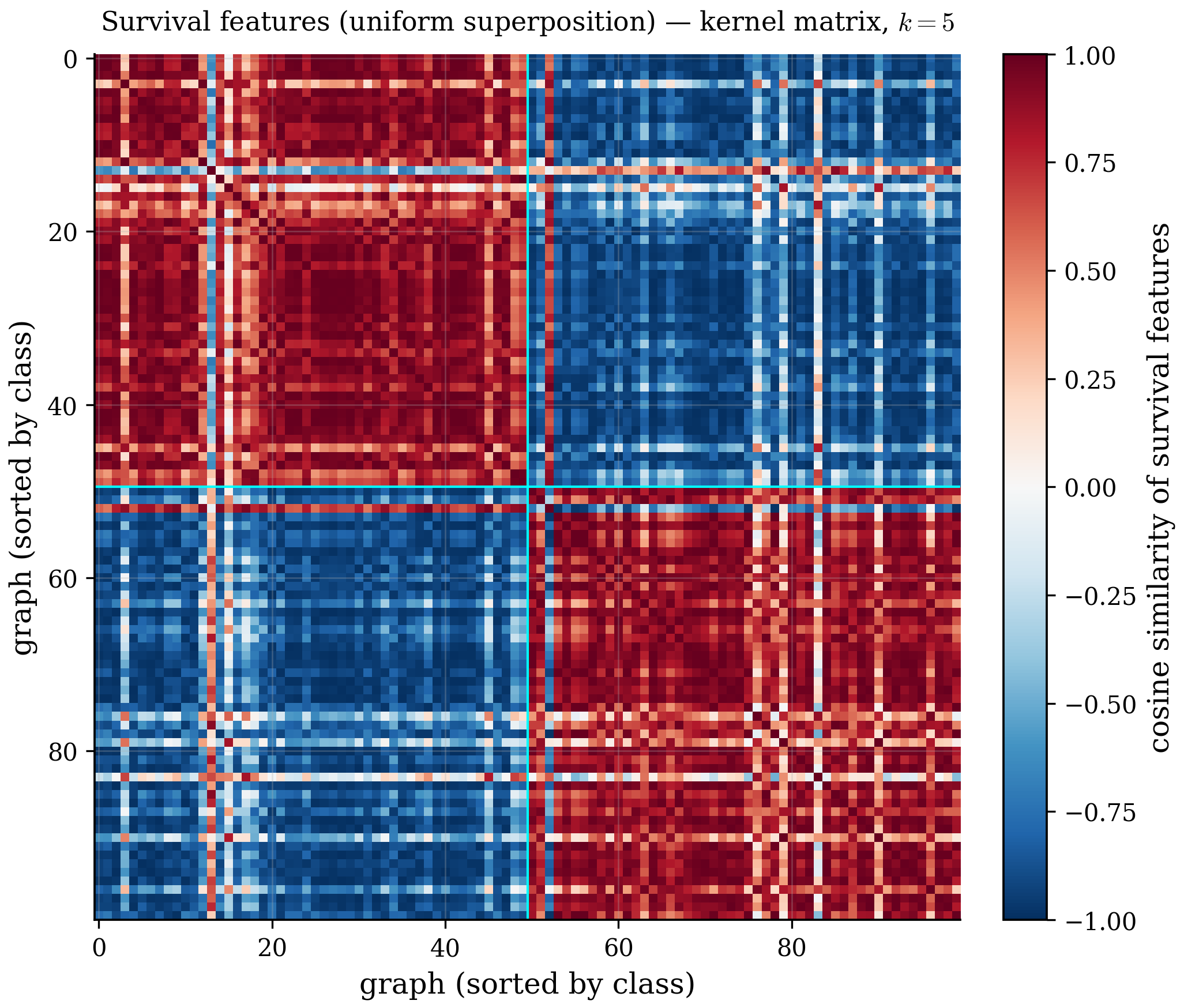}
    \caption{$k=5$}\label{fig:surv-k5}
  \end{subfigure}\hfill
  \begin{subfigure}[t]{0.32\textwidth}
    \includegraphics[width=\linewidth]{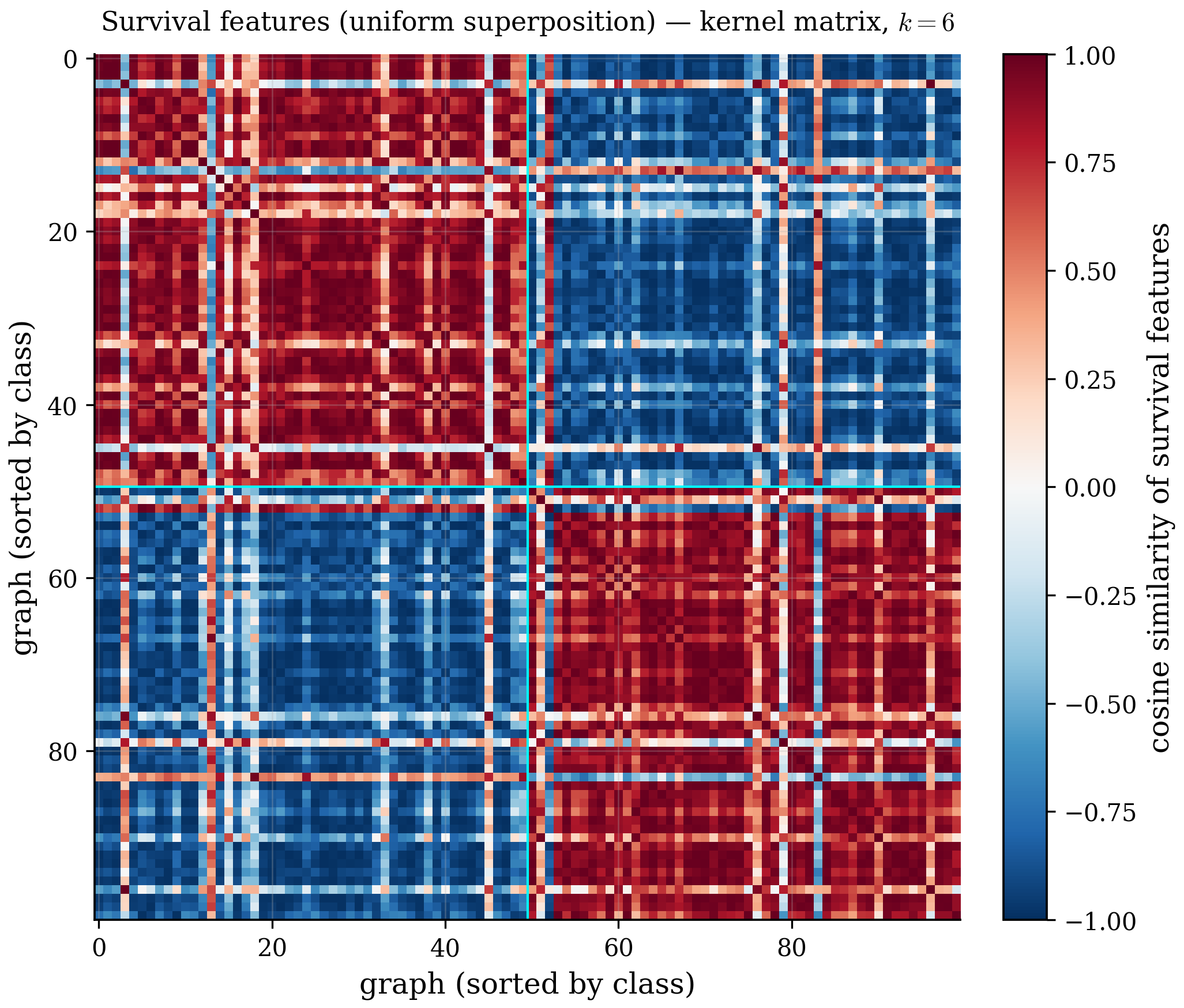}
    \caption{$k=6$}\label{fig:surv-k6}
  \end{subfigure}
 
  \medskip
  \begin{subfigure}[t]{0.32\textwidth}
    \includegraphics[width=\linewidth]{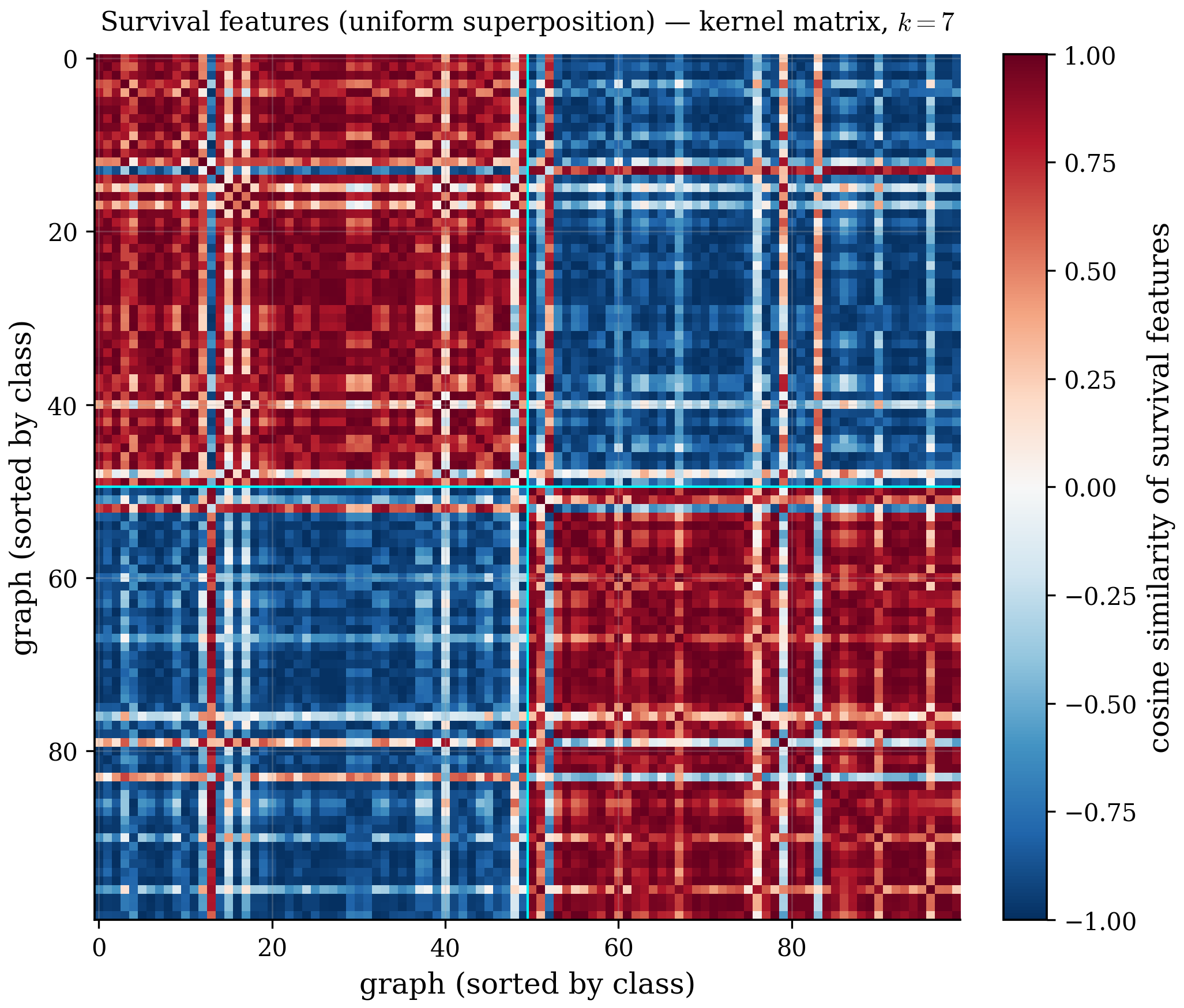}
    \caption{$k=7$}\label{fig:surv-k7}
  \end{subfigure}\hfill
  \begin{subfigure}[t]{0.32\textwidth}
    \includegraphics[width=\linewidth]{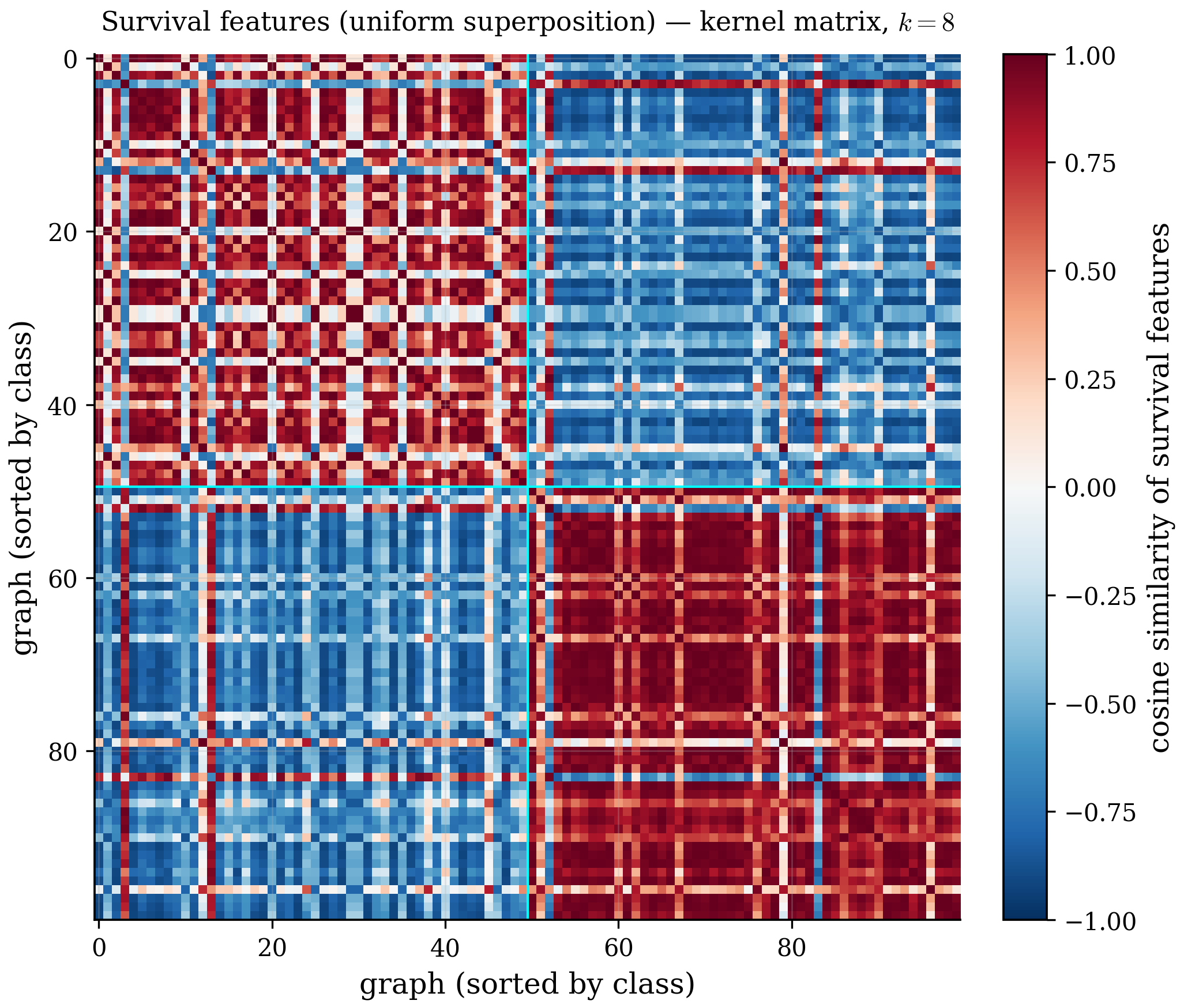}
    \caption{$k=8$}\label{fig:surv-k8}
  \end{subfigure}\hfill
  \begin{subfigure}[t]{0.32\textwidth}
    \includegraphics[width=\linewidth]{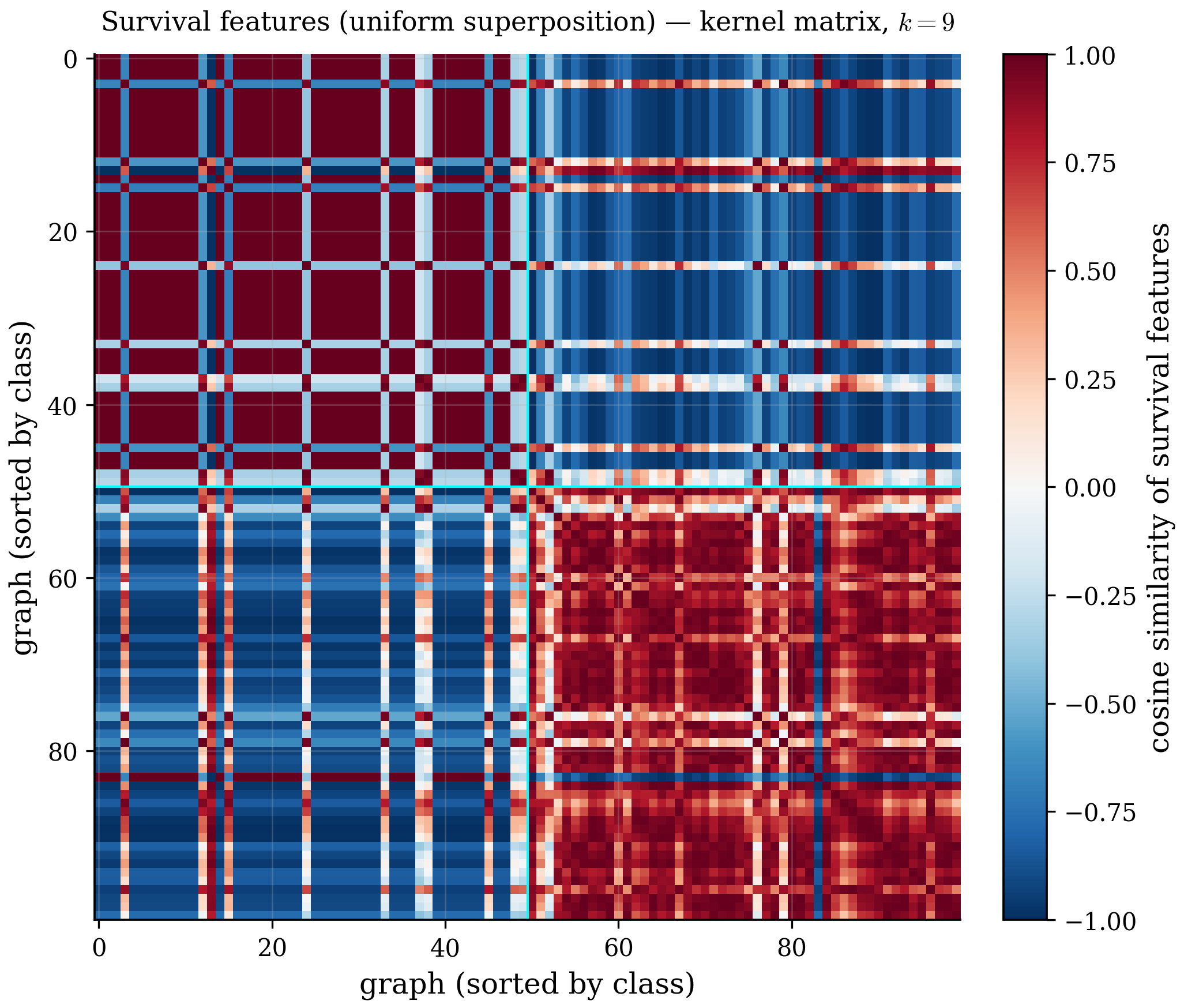}
    \caption{$k=9$}\label{fig:surv-k9}
  \end{subfigure}
 
  \caption{\justifying Survival-feature cosine-similarity matrices for Laplacian
           dimensions $k=1,\dots,9$, each evaluated on $G(50, p)$ with $25$ graphs per class, for class $1$: $p_0=0.5$ and class $2$: $p_1=0.52$. Features are
           the standardised $[\,\mathrm{Re}\,a(t),\ \mathrm{Im}\,a(t),\ |a(t)|^2\,]$
           survival series; rows/columns are sorted by class, cyan cross marks
           the class boundary.}
  \label{fig:survival-grid}
\end{figure}

This appendix complements the representative examples presented in the main text by reporting the complete set of similarity matrices for simplicial dimensions $k=1,\ldots,9$. Figure~\ref{fig:fidelity-grid} shows the pairwise fidelity kernels corresponding to the implicit quantum representations, while Figure~\ref{fig:survival-grid} presents the cosine similarity matrices of the explicit measurement-derived feature vectors. The two figures illustrate how the geometry of both representations changes across simplicial dimensions and further support the observation that the explicit features preserve much of the class structure captured by the underlying quantum states.

\section{Additional data on spectral filters}
\begin{figure}
    \centering
    \includegraphics[width=0.89\linewidth]{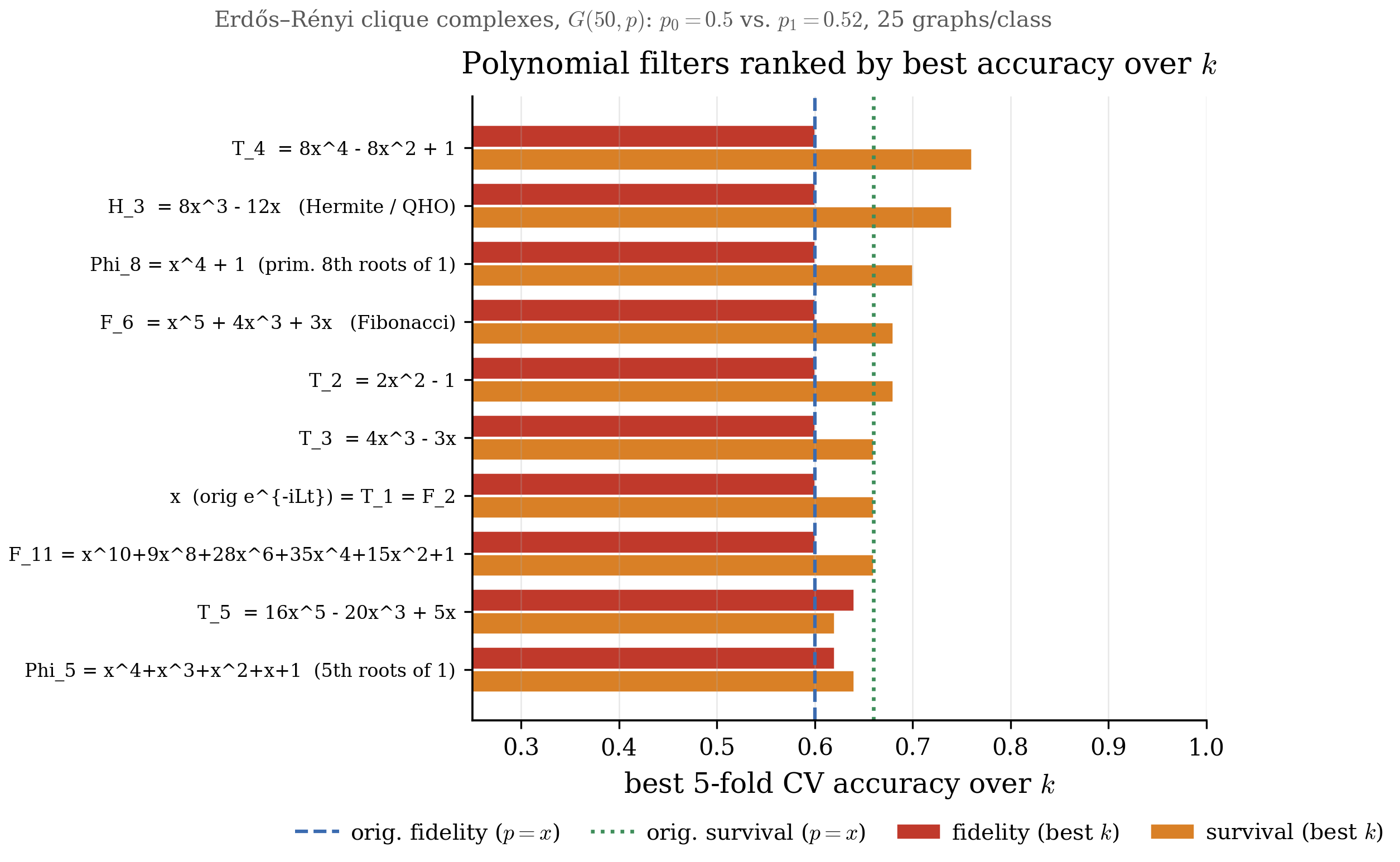}
    \caption{\justifying Performance comparison of the polynomial spectral filters evaluated in the QSVT experiments. Bars indicate the relative improvement or degradation in classification accuracy obtained with each family of polynomial filters compared to the baseline topology-driven quantum evolution. The results illustrate the impact of the chosen spectral transformation on the quality of the induced quantum representations.}
    \label{fig:Specific_Poly}
\end{figure}

This appendix provides a detailed comparison of the polynomial spectral filters considered in the QSVT experiments. While the main text focuses on the best-performing filter families, Figure~\ref{fig:Specific_Poly} reports the performance of all polynomial classes evaluated in this work. The results illustrate how different spectral transformations of the combinatorial Laplacian influence the quality of the resulting quantum representations, highlighting the sensitivity of QTDE to the choice of spectral filter and motivating the use of task-dependent polynomial transformations.

\label{sec:specificpoly}

\end{document}